\documentclass[twocolumn,pre,showpacs,amsmath,amssymb,nobibnotes,nofootinbib,superscriptaddress]{revtex4}
\usepackage{natbib}
\usepackage{amsmath}
\usepackage{dcolumn}
\usepackage{wrapfig}
\usepackage{graphicx}
\usepackage{color}
\usepackage{graphics}
\usepackage{epsfig}
\usepackage{units}
\usepackage{dsfont}
\usepackage{wasysym}

\parskip=\medskipamount

\newcommand{\fig}[1]{Fig.~\ref{#1}}

\newcommand{\be}{\begin{equation}}
\newcommand{\ee}{\end{equation}}

\begin{document}

\title{Dynamic message-passing equations for models with unidirectional dynamics}

\author{Andrey Y. Lokhov}
\email{andrey.lokhov@lptms.u-psud.fr}
\affiliation{Universit\'e Paris-Sud/CNRS, LPTMS, UMR8626, B\^at. 100, 91405 Orsay, France}
\author{Marc M\'ezard}
\affiliation{Universit\'e Paris-Sud/CNRS, LPTMS, UMR8626, B\^at. 100, 91405 Orsay, France} 
\affiliation{Ecole normale sup\'erieure - PSL Research University, 45 rue d'Ulm, 75005 Paris, France}
\author{Lenka Zdeborov\'a}
\affiliation{Institut de Physique Th\'eorique, IPhT, CEA Saclay and CNRS URA 2306, 91191 Gif-sur-Yvette, France}
\date{\today}

\begin{abstract}
Understanding and quantifying the dynamics of disordered out-of-equilibrium models is an important problem in many branches of science. Using the dynamic cavity method on time trajectories, we construct a general procedure for deriving the dynamic message-passing equations for a large class of models with unidirectional dynamics, which includes the zero-temperature random field Ising model, the susceptible-infected-recovered model, and rumor spreading models. We show that unidirectionality of the dynamics is the key ingredient that makes the problem solvable. These equations are applicable to single instances of the corresponding problems with arbitrary initial conditions, and are asymptotically exact for problems defined on locally tree-like graphs. When applied to real-world networks, they generically provide a good analytic approximation of the real dynamics.
\end{abstract}
\maketitle

\section{Introduction}

Over the past decade, there has been a growing interest in building analytical tools for the study of  out-of-equilibrium dynamics in disordered  problems defined on heterogeneous networks. A particular attention has been devoted to the study of cascading and avalanche processes, in the cases where the dynamics is not a relaxation dynamics related to a Hamiltonian, but instead is characterized by a set of stochastic transition rules. Examples of such processes include epidemic spreading \cite{Murray89,anderson1991infectious,Hethcote00,BoccalettiaLatora06}, propagation of information and innovations in social media \cite{Rogers2010,Chakrabarti2008,Llas2003,Guardiola2002,Strang1998}, dynamics of magnetic and glassy systems \cite{Dhar1997,OhtaSasa2010}, communication protocols, such as gossip algorithms and peer-to-peer file sharing on computer networks \cite{Demers1987,Vogels2003}, activation cascades in biological and neural networks \cite{o2013spreading,kholodenko2012computational}, and news updates in financial markets \cite{Kimmel2004, Kosfeld1998}. A common property shared by these processes is the unidirectional nature of the corresponding dynamics: once an elementary constituent of the system under the influence of its neighbors undergoes a transition to a certain state, it can never return to the previous one.

Although the properties of diluted disordered systems have been intensively investigated over the past several years, there is still no well-established tractable method for solving the corresponding dynamics in the general case. One category of problems that has recently attracted a lot of attention is the case of 
out-of-equilibrium dynamic processes on sparse graphs \cite{Hatchett2004,Mimura2009,Neri2009,Kanoria2011}. Methods which are developed in this context can also be used as sophisticated mean-field-type approximations for problems defined on general graphs.
The generating functional analysis techniques \cite{Dominicis78}, the dynamical replica analysis \cite{Hatchett2005,mozeika2009dynamical} and the cavity method \cite{MezardParisi01,MezardMontanari2009} have been recently used for the construction of a general approach in terms of time-trajectories of variables.
However, the general dynamics remains intractable in this formalism except for only a few time steps: the solution of the corresponding equations takes in general a number of operations that grows exponentially with the duration of the process one wants to study. In a few special cases, some progress has been recently made by a number of authors who were able to write, using cavity-like arguments, tractable asymptotically-exact mean-field dynamic equations for several models defined on locally tree-like graphs, such as the zero-temperature random field Ising model (RFIM) \cite{OhtaSasa2010}, the susceptible-infected-recovered (SIR) model \cite{KarrerNewman10b,Volz08,Miller11,lucas2012exact,PatientZero2013}, and the threshold models \cite{Altarelli2013,shrestha2014message}. All these models share a common property: they describe a unidirectional dynamics involving one transition to the active state; the derivation of the corresponding equations is typically based on identifying correct dynamic variables that are required to obtain the closed-form expressions. These examples lead to the hypothesis that the microscopic irreversibility of the dynamics is a key property that makes it possible to derive such equations \cite{lucas2012exact,altarelli2013optimizing}. However, in general it is very difficult to guess the right dynamic variables that should be used in the dynamic equations for more complicated models, involving a larger number of states and several non-trivial transitions. Probably, the simplest model of this kind is the so-called rumor spreading model \cite{DaleyD.J.&Kendall1964, DaleyD.J.&Kendall1965,MakiD.P.&Thompson1973}, which is a three-state dynamic model with two neighbors-dependent transitions.

In this paper, we develop a systematic procedure for deriving the dynamic message-passing (DMP) equations for general models with unidirectional dynamics and arbitrary number of states. They allow one to estimate the marginal probabilities of each variable at each time on a given network of contacts, using a number of operations that is polynomial both in the size of the network and in the duration of the dynamic process. These equations are applicable to single-instance problems with arbitrary initial conditions, they are asymptotically exact on locally tree-like networks, and typically provide a good approximation for real-world networks. The DMP equations are derived using the cavity method, also known in different fields as the belief propagation (BP), or the sum-product algorithm \cite{YedidiaFreeman2003,MezardMontanari2009}, starting from a BP equation that takes as variables the time trajectories of nodes. Despite the similarity with the BP equations that need to be iterated until convergence, the iteration time in the DMP equations corresponds to the physical time. We show that the unidirectional nature of the dynamics is indeed a crucial element that makes the problem solvable. More precisely, the time trajectories in these models can be fully parametrized with only a few flipping times, leading to a significant simplification of the corresponding dynamic BP formulation. As a result, these BP equations can be rewritten in terms of closed-form DMP equations with a computational complexity which turns out to be reduced from an exponential in the duration of the process to a polynomial. This simplification occurs thanks to the use of dynamic variables that appear naturally to be the weighted sums of messages of the BP equations on trajectories.

The structure of the paper is as follows. In section \ref{sec:DBP} we present a systematic framework for treating dynamic problems on locally tree-like graphs, writing a general dynamic belief propagation equation for node trajectories in time. Then, in section \ref{sec:Irreversible} we introduce an important class of considered models with unidirectional dynamics. In the next sections \ref{sec:K2}-\ref{sec:K4} we construct a general procedure for the derivation of DMP equations for these models, thus obtaining the single-instance form of the mean-field equations that for some cases have already appeared in the literature, and obtaining new equations for more complicated models. Finally, we provide supporting numerical results and discuss the possible applications of our approach.

\section{Dynamic belief propagation}
\label{sec:DBP}

Belief propagation, or the cavity method, is an iterative method that allows one to estimate efficiently marginal probability distributions in graphical models. It has been proven to be very successful in some applications, e.g. error-correcting codes \cite{Gallager62}, Bayesian networks \cite{Pearl88}, and optimization problems \cite{MezardParisi02}. BP makes use of the assumption that the marginal probabilities (called messages) defined on an auxiliary cavity graph (a graph with a removed node) are uncorrelated. This assumption is obviously exact if the underlying network is a tree, in other cases it is an approximation of the mean-field type (for more details, see \cite{MezardMontanari2009}). The solution to the BP equations can often be obtained by iterating the equations until convergence. A formulation of this algorithm for static problems is given for consistency in Appendix \ref{app:BPstatic}.

Our motivation to seek a generalization of the cavity method for dynamic problems has been inspired by the success of belief propagation in static problems. The main idea behind the dynamic belief propagation is to write usual cavity equations using the time trajectories of nodes as variables. This idea has been exploited in a number of previous works on the dynamics of disordered systems \cite{Kanoria2011,Neri2009,Hatchett2004,Mimura2009}. 

Consider a graph $G=(V,E)$, defined by a vertex set $V$ and a set of edges $E$. In the dynamic setting, each vertex $i \in V$ is characterized by a variable, taking the value $\sigma_{i}^{t}$ at time $t$. We assume that the set of possible values of  $\sigma_{i}^{t}$  is of size $K$. We consider a generic dynamic process defined in a discrete-time parallel dynamics and described by a local transition probability $w_{i}(\sigma_{i}^{t+1}\mid\{\sigma_{j}^{t}\}_{j\in\partial i})$ that a node $i$ takes value $\sigma_{i}^{t+1}$ at time $t+1$ given the values $\{\sigma_{j}^{t}\}$ of its neighbors at time $t$. If we denote by $\vec{\sigma}_{i}=(\sigma_{i}^{0},\ldots,\sigma_{i}^{T})$ the trajectory of variable $i$ at times $t=0,\ldots,T$, where $T$ is the stopping time, the joint probability distribution of the trajectories $P(\{\vec{\sigma}_{i}\}_{i\in V})$ can be written as follows:
\begin{equation}
P(\{\vec{\sigma}_{i}\}_{i\in V})=\prod_{i\in V}\prod_{t=0}^{T-1}w_{i}(\sigma_{i}^{t+1}\mid\{\sigma_{j}^{t}\}_{j\in\partial i})P_{0},
\label{dyndef}
\end{equation}
where  $P_{0} \equiv P(\{\sigma_{i}^{0}\}_{i\in V})$ is the distribution of variables at initial time.

\begin{figure}[!ht]
\centering
\includegraphics[scale=0.7]{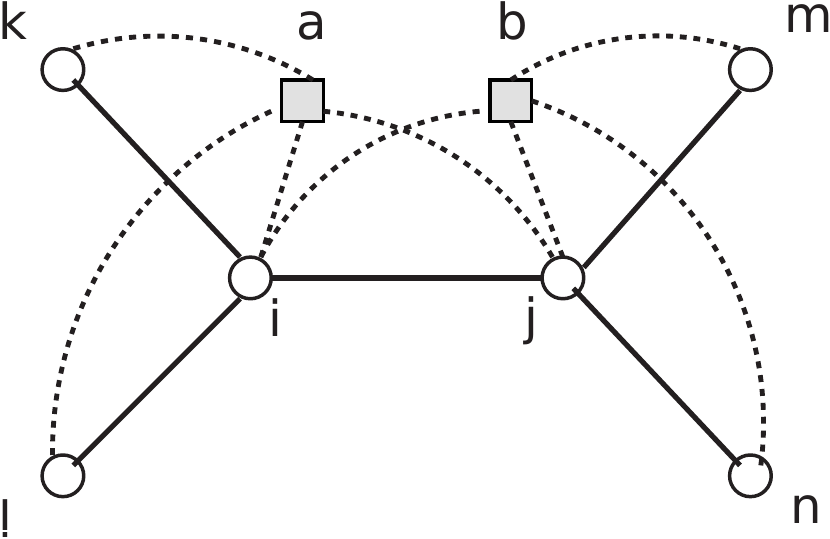}
\caption{An example of a factor graph of the graphical model at two nearest times described by the joint probability distribution $P(\{\vec{\sigma}_{i}\}_{i\in V})$. The check node $a$ represents interaction between the variable $\sigma_{i}^{t+1}$ and the variables $\{\sigma_{j}^{t}\}_{j\in\partial i}$ at a previous time step. This factor graphs is characterized by systematic short loops.}
\vspace{-0.1cm}
\label{fig:1}
\end{figure}

It is a well-known fact that BP equations are exact for static graphical models when the factor graph is a tree. However, when we consider the factor graph (graph involving check nodes that represent local interactions between variables) of the model defined in \eqref{dyndef}, in which the variables are time trajectories $\vec{\sigma}_{i}$, it turns out that the factor graph contains many loops, even in the case where $G$ is a tree, see \fig{fig:1}.

\begin{figure}[!ht]
\centering
\includegraphics[scale=0.7]{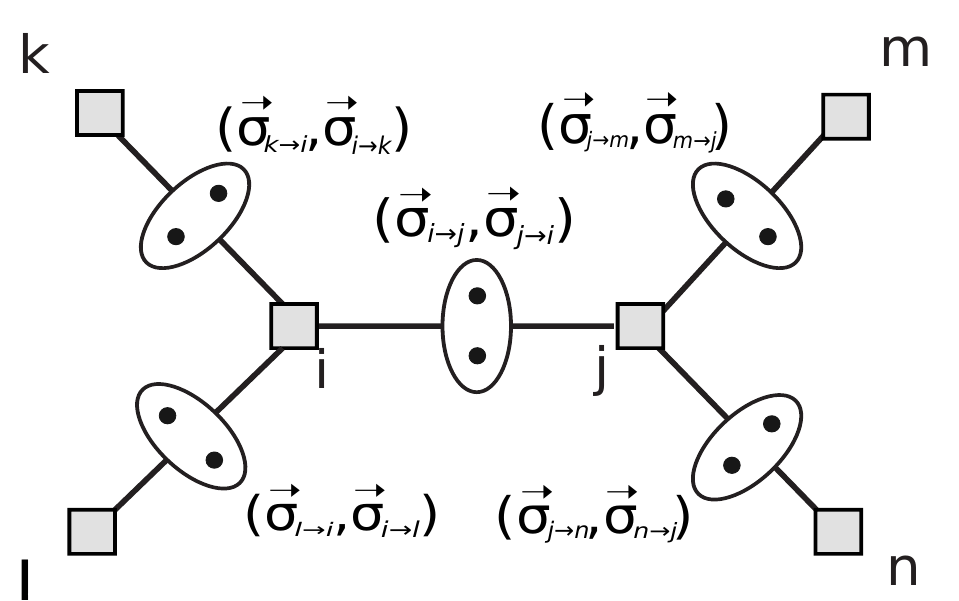}
\caption{An example of a factor graph of the graphical model at all times described by the joint probability distribution $P(\{\vec{\sigma}_{i\rightarrow j}, \vec{\sigma}_{j\rightarrow i}\}_{(i,j)\in E})$. The check node $i$ represents interaction between trajectories $\vec{\sigma}_{i}$ and $\{\vec{\sigma}_{j}\}_{j\in\partial i}$. This factor graph is characterized by the underlying tree structure if the original graph is a tree.}
\vspace{-0.1cm}
\label{fig:2}
\end{figure}

A way to fix this problem consists in exploiting the duality between variables and interactions by putting the variables on the edges. To this purpose we introduce a different representation of the problem, that uses auxiliary variables (time-trajectories) $\vec{\sigma}_{i\rightarrow j}$ on each directed edge $(i,j)\in E$. For a given $i$, all the variables $\vec{\sigma}_{i\rightarrow j}$ are supposed to be copies of the original $\vec{\sigma}_{i}$. They should thus be all equal, and we implement this by adding for each $i$ an additional constraint $\vec{\sigma}_{i\rightarrow j}=\vec{\sigma}_{i\rightarrow k}$ for all $j,k\in \partial i$. The joint probability distribution \eqref{dyndef} of time trajectories can hence be written in terms of these new variables:
\begin{align}
\notag
&P(\{  \vec{\sigma}_{i\rightarrow j}, \vec{\sigma}_{j\rightarrow i}\}_{(i,j)\in E})
\\
&=\prod_{i\in V}\prod_{t=0}^{T-1}
 \Bigg[
w_{i}(\sigma_{i\rightarrow l}^{t+1}\mid\{\sigma_{k\rightarrow i}^{t}\}_{k\in\partial i})
\prod_{k\in\partial i\backslash l}\delta_{\sigma^{t}_{i\rightarrow l},\sigma^{t}_{i\rightarrow k}}
\Bigg]
P_{0},
\label{eq:Joint_proba_distri_pairs}
\end{align}
where $l$ is any of the variables influenced directly by $i$, and $k\in \partial i\backslash l$ means the set of nodes neighboring node $i$, excluding $l$. This new form of the probability distribution is described by a factor graph which is very closely related to $G$ : the new variables $\vec{\sigma}_{i\rightarrow j},\vec{\sigma}_{j\rightarrow i}$ live on each edge $(ij)\in E$, and there is a function node (interaction) associated with every vertex $i\in V$. If the original graph $G$ is a tree (respectively, is locally tree-like), the factor graph is a tree (respectively, is locally tree like), see \fig{fig:2}. This crucial property allows to use the BP method in terms of this new description for studying the dynamics, with the guarantee that the resulting equations are exact if $G$ is a tree.

Let us now write the BP equations. Using the fact that $\vec{\sigma}_{i\rightarrow j}=\vec{\sigma}_{i\rightarrow k}$ for all $j,k\in \partial i$, it is convenient to rename the variables $\{\vec{\sigma}_{i\rightarrow j}, \vec{\sigma}_{j\rightarrow i}\}_{(i,j)\in E}$ to $\{\vec{\sigma}_{i}, \vec{\sigma}_{j}\}_{(i,j)\in E}$. The BP equations for the joint probability distribution \eqref{eq:Joint_proba_distri_pairs} in terms of conditional messages $m^{i\rightarrow j}(\vec{\sigma}_{i} \mid \vec{\sigma}_{j})$ read:
\begin{align}
\notag
m^{i\rightarrow j}(\vec{\sigma}_{i}\mid\vec{\sigma}_{j})=
\hspace{-0.34cm}
& \sum_{\{\vec{\sigma}_{k}\}_{k\in\partial i\backslash j}}
\hspace{-0.14cm}
\left[
\prod_{t=0}^{T-1}w_{i}(\sigma_{i}^{t+1}\mid\{\sigma_{k}^{t}\}_{k\in\partial i\backslash j},\sigma_{j}^{t})
\right]
\\
& \times P(\{\sigma_{i}^{0}\}_{i\in V})
\hspace{-0.23cm}
\prod_{k\in\partial i\backslash j}m^{k\rightarrow i}(\vec{\sigma}_{k}\mid\vec{\sigma}_{i}).
\label{eq:DBP_conditional}
\end{align}
Notice that, in general, there appears a normalization constant in front of the BP equation. In our case, thanks to the Markov property of the dynamics, we can explicitly compute the normalization constant, see Appendix~\ref{app:DBPdetails} for details.

The message $m^{i\rightarrow j}(\vec{\sigma}_{i}\mid\vec{\sigma}_{j})$ has the meaning of the probability for the trajectory $\vec{\sigma}_{i}$ given the trajectory $\vec{\sigma}_{j}$ in the transformed cavity graph, where the factor node $j$ has been removed. We denote the dynamics in the corresponding cavity graph as $D_{j}$. The equation \eqref{eq:DBP_conditional} can be iterated until convergence, and the corresponding marginal probability of a time trajectory $\vec{\sigma}_{i}$ will be given by
\begin{align}
\notag
m^{i}(\vec{\sigma}_{i})=\hspace{-0.14cm}\sum_{\{\vec{\sigma}_{k}\}_{k\in\partial i}}&\left[\prod_{t=0}^{T-1}w_{i}(\sigma_{i}^{t+1}\mid\{\sigma_{k}^{t}\}_{k\in\partial i})\right]
\\
&\times P(\{\sigma_{i}^{0}\}_{i\in V})\prod_{k\in\partial i}m^{k\rightarrow i}(\vec{\sigma}_{k}\mid\vec{\sigma}_{i}).
\label{eq:DBP_marginal_conditional}
\end{align}
Note that in the general case, it takes an exponential number of operations in the duration of the process to solve the equations \eqref{eq:DBP_conditional} and \eqref{eq:DBP_marginal_conditional}, since each message has $K^{T}$ components, where $K$ is the number of values that each variable $\sigma_{i}^{t}$ may take, and the sum in \eqref{eq:DBP_conditional} is performed over $K^{T(d_{i}-1)}$ variables for each node $i$, with $d_{i}$ being the number of neighbors of $i$. However, a crucial simplification occurs for the models with unidirectional dynamics, introduced in the next section.

\section{Models with unidirectional dynamics}
\label{sec:Irreversible}

Let us assume that in the expression for the transition probability $w_{i}(\sigma_{i}^{t+1}\mid\{\sigma_{j}^{t}\}_{j\in\partial i})$, the value $\sigma_{i}^{t}$ takes one of the $K$ ordered discrete values that we denote $\Omega_{1},\Omega_{2},\ldots,\Omega_{K}$. We call the dynamic process unidirectional if the node can change its state only in a directed and irreversible way, $\Omega_{1} \rightarrow \Omega_{2} \rightarrow \ldots \rightarrow \Omega_{K}$, and the transition to one of the previous states is forbidden by the dynamic rules.

Among unidirectional processes with $K=2$ states ($\sigma_{i}^{t}$ can take one of the two values $-1\equiv \downarrow$ or $1\equiv \uparrow$), one can mention the zero-temperature random field Ising model (zero-temperature RFIM) with homogeneous initial condition, considered in \cite{OhtaSasa2010}: $\prod_{i\in V}\delta_{\sigma_{i}^{0},-1}=(\downarrow\downarrow\cdots\downarrow)$. Each spin can flip only if the local field created by its neighbors is positive (precise definitions will be given in the next section). Once being flipped, the spins in this avalanche dynamics remain in the position $\uparrow$ for all times, since the local field is a monotone non-decreasing function of time. Therefore, this system has a unidirectional dynamics with two ordered states, $\downarrow$ and $\uparrow$. The model can be generalized to any initial condition if the backward transition from $\uparrow$ to $\downarrow$ is explicitly forbidden by the dynamic rules.
The situation is different from the standard Glauber dynamics of Ising model with non-zero temperature \cite{glauber1963time}, or the majority dynamics of voters, that switch to one of the alternative opinions according to the majority of their neighbors \cite{Kanoria2011}, where each variable is free to flip an arbitrary number of times. Note that the linear threshold model with random thresholds studied in \cite{Altarelli2013,altarelli2013optimizing} is equivalent to this formulation of the RFIM.

Another example of model with two states and unidirectional dynamics is given by the susceptible-infected (SI) model, in which the node can be in either of two states: susceptible ($S$), or infected ($I$). The propagation of infection on this model occurs due to the pairwise interactions between individuals: for instance, the $S$ individual can be infected by one of its $I$ neighbors at each time step, and then remains infected forever. If there exists a recovery mechanism that allows an infected individual to become susceptible again after some time (it corresponds to the so-called susceptible-infected-susceptible, or SIS model, that is used to model the behavior of endemic diseases), the model does not belong to the class of unidirectional models anymore.

Unidirectional dynamic processes with $K=3$ states include the susceptible-infected-recovered (SIR) model, an extension of the SI model that is widely used to model epidemic spreading. In this model, the infected node can switch to a recovered ($R$) state with a certain probability at each time step, leading to a depletion of infected agents. Another well known model with unidirectional dynamics and three states is given by the so-called {\it rumor spreading}, or ignorant-spreader-stifler (ISS) model \cite{BoccalettiaLatora06}, which describes the propagation of information by spreaders to ignorants that are unaware of rumor, and takes into account the possibility that the spreader can become uninterested in the rumor under the influence of its neighbors. The precise formulation of these models will be given in the section \ref{sec:K3}. 

In what follows, we discuss the dynamic message-passing equations for the models listed above, illustrating the general method to derive such equations for other models with arbitrary $K$. Typically, these equations would allow us to answer the following question: what is the probability that a certain node $i$ is in a certain state $\Omega_{a}$ at time $t$? For some of these models, the equivalent equations have already appeared in the literature, however, in a form averaged over the ensemble of random graphs and/or over the initial conditions, and not suitable to the algorithmic purposes for single-instance problems. For others, the DMP equations have never been stated previously; the ``naive'' mean-field equations that exist for all models are derived under assumptions of homogeneity of transmission probabilities and complete-mixing hypothesis (ignoring the actual topology of the interaction network and assuming that each pair of nodes may interact) that are obviously unrealistic. For each model considered in the following, we will discuss the relation of our DMP equations to those existing in the literature, if any.     

\section{Unidirectional models with $K=2$ states}
\label{sec:K2}

\subsection{Zero-temperature RFIM}

The RFIM Hamiltonian reads
\begin{equation}
H=-\sum_{(ij)}J_{ij}\sigma_{i}\sigma_{j}-\sum_{i}(h+h_{i})\sigma_{i},
\end{equation}
where $J_{ij}$ is a non-negative interaction between spins $i$ and $j$, $h$ is an external uniform magnetic field, and $h_{i}$ is a random magnetic field on site $i$, extracted from some probability distribution.
At zero temperature spin $i$ tends to be aligned with its local magnetic field
\begin{equation}
\Delta^{t}_{i}=h+h_{i}+\sum_{k\in\partial i}J_{ki}\sigma^{t}_{k}.
\label{eq:RFIM_local_field}
\end{equation}

Consider an initial condition in the form $P(\{\sigma_{i}^{0}\}_{i\in V})=\prod_{i\in V}\delta_{\sigma_{i}^{0},-1}$ (as mentioned in the previous section, one could choose any initial condition, provided the dynamics is such that the transition from $\uparrow$ to $\downarrow$ is forbidden). Define the zero temperature stochastic dynamics respecting the following property: spin $\sigma_{i}=-1$ with a positive local field $\Delta_{i}$ flips with rate $1/\tau$, and does not flip otherwise. Each spin flips at most only once, so the trajectory $\vec{\sigma}_{i}$ has a typical form $\left\vert\downarrow_{_{0}}\downarrow\downarrow\downarrow\downarrow\downarrow\downarrow
\downarrow\uparrow_{_{\tau_{i}}}\uparrow\uparrow\uparrow\uparrow\uparrow\uparrow\uparrow_{_{T}}\right\rangle$ and is in the one-to-one correspondence with the flipping time $\tau_{i}$ ($\tau_{i}$ is the first time for which $\sigma_{i}=1$). If the spin does not flip for all the times $0,\ldots,T-1$, then by definition we set $\tau_{i}=T$ ($T$ is the stopping time, i.e. the condition $\tau_{i}=T$ summarizes all the events that happen after the time $T$).

Using the representation in terms of flipping times, Eq. (\ref{eq:DBP_conditional}) for $\tau_{i} < T$ can be expressed as follows:
\begin{equation}
m^{i\rightarrow j}(\tau_{i}\mid\tau_{j})=
\hspace{-0.25cm}
\sum_{\{\tau_{k}\}_{k\in\partial i\backslash j}}
\hspace{-0.34cm}
W_{RFIM}
\hspace{-0.15cm}
\prod_{k\in\partial i\backslash j}m^{k\rightarrow i}(\tau_{k}\mid\tau_{i}),
\label{eq:message_RFIM_one}
\end{equation}
where
\begin{equation}
W_{RFIM}=\prod_{t'=0}^{\tau_{i}-2}\left(1-\frac{1}{\tau}\mathds{1}[\Delta_i^{t'}>0]\right)\frac{1}{\tau}\mathds{1}[\Delta_i^{\tau_{i}-1}>0].
\end{equation}
Here and in what follows we use a convention
\begin{equation}
\prod_{t=a}^{a-\epsilon}(\ldots) \equiv 1
\label{eq:convention}
\end{equation}
for any fixed $a$ and $\epsilon > 0$. Using the fact that messages are properly normalized, we choose
\begin{equation}
m^{i\rightarrow j}(T \mid\tau_{j}) = 1 - \sum_{\tau_{i}=1}^{T-1}m^{i\rightarrow j}(\tau_{i}\mid\tau_{j}).
\end{equation}

From the conditional messages $m^{i\rightarrow j}(\tau_{i}\mid\tau_{j})$, we can define two quantities:
\begin{align}
p^{i\rightarrow j}(t)=
\sum_{\tau_{i} > t}m^{i\rightarrow j}(\tau_{i}\mid T), \label{eq:p_definition}
\\
q^{i\rightarrow j}(t)=
\sum_{\tau_{i} \leq t}m^{i\rightarrow j}(\tau_{i}\mid T).
\label{eq:q_definition}
\end{align}
These quantities characterize the marginals of the zero-temperature RFIM in the cavity dynamics $\mathcal{D}_{j}$, in which $\sigma_{j}^{t}=-1$ for every $t$ and never flips, even if $\Delta_{j}>0$. In this dynamics, $p^{i\rightarrow j}(t)$ is the probability that spin $i$ stays in the state $\sigma_{i}^{t}=-1$ at time $t$, and $q^{i\rightarrow j}(t)$ is defined to be the probability that spin $i$ has already flipped, and hence $\sigma_{i}^{t}=1$.

Of course, we will be ultimately interested in writing a closed equation for the marginals in the original dynamics, defined as:
\begin{align}
p^{i}(t)=
\sum_{\tau_{i} > t}m^{i}(\tau_{i}), \label{eq:p_marginal_definition}
\\
q^{i}(t)=1-p^{i}(t),
\label{eq:q_marginal_definition} 
\end{align}
where $m^{i}(\tau_{i})$ are the marginal probabilities of trajectories \eqref{eq:DBP_marginal_conditional} in the original dynamics, following the same equation as \eqref{eq:message_RFIM_one}, but with $\partial i \backslash j$ replaced by $\partial i$, in the same way as \eqref{eq:DBP_marginal_conditional} is related to \eqref{eq:DBP_conditional}.
The DMP equations for this model can be derived, starting from the equation (\ref{eq:message_RFIM_one}), and using the definitions \eqref{eq:p_definition}, \eqref{eq:q_definition}, as well as elementary properties of the messages, such as normalization and causality constraints. For details of the derivation see Appendix \ref{app:RFIMdetails}. After some algebra, the resulting DMP equations can be shown to take the following form in discretized time notations:
\begin{align}
\notag
q^{i\rightarrow j}(t+&1)= \left(1-\frac{1}{\tau}\right) q^{i\rightarrow j}(t)
\\
\notag
&+\frac{1}{\tau}
\sum_{\{\sigma_{k}\}{}_{k\in\partial i\backslash j}}
\hspace{-0.34cm}
\mathds{1}\left[h+h_{i}+
\hspace{-0.25cm}
\sum_{k\in\partial i\backslash j}
\hspace{-0.25cm}
J_{ki}\sigma_{k} -J_{ji}>0\right]
\\
&\times\prod_{k\in\partial i\backslash j:\sigma_{k}=+1}
\hspace{-0.45cm}
q^{k\rightarrow i}(t)
\hspace{-0.45cm} \prod_{k\in\partial i\backslash j:\sigma_{k}=-1}
\hspace{-0.57cm}
\left[1-q^{k\rightarrow i}(t)\right].
\label{eq:Exact_dynamic_RFIM}
\end{align}
Therefore, the marginal probability for spin $i$ to be in the state $+1$ at time $t+1$ is given by $q^{i}(t+1)$, which can be computed according to the following expression:
\begin{align}
\notag
q^{i}(t+&1) = \left( 1 - \frac{1}{\tau} \right) q^{i}(t)
\\
\notag
&+\frac{1}{\tau}
\sum_{\{\sigma_{k}\}{}_{k\in\partial i}}
\hspace{-0.16cm}
\mathds{1}\left[h+h_{i}+
\sum_{k\in\partial i}
J_{ki}\sigma_{k}>0\right]
\\
&\times\prod_{k\in\partial i:\sigma_{k}=+1}
\hspace{-0.39cm}
q^{k\rightarrow i}(t)
\hspace{-0.39cm} \prod_{k\in\partial i:\sigma_{k}=-1}
\hspace{-0.39cm}
\left[1-q^{k\rightarrow i}(t)\right].
\label{eq:RFIM_marginal}
\end{align}
The probability that spin $i$ is still in the state $-1$ at time $t+1$ is then given by $p^{i}(t+1)=1-q^{i}(t+1)$. Note that the DMP equations \eqref{eq:Exact_dynamic_RFIM} and \eqref{eq:RFIM_marginal} can now be run in the real time, starting with initial conditions $q^{i}(0)=q^{i\rightarrow j}(0)=0$ for each node $i$ and $j$; these equations have a closed self-consistent form, so we no longer need to compute the messages using \eqref{eq:message_RFIM_one}. Note that the computational complexity of the DMP equations for the zero-temperature RFIM is reduced from exponential to linear in time; in the most straightforward implementation, the computation complexity of \eqref{eq:Exact_dynamic_RFIM} and \eqref{eq:RFIM_marginal} is $\mathcal{O}(N 2^{c} t)$, where $c$ is the average degree of the graph.

The averaged form of the DMP equations has been first derived in \cite{OhtaSasa2010} using a cavity-like argument for the dynamic variables $q^{i\rightarrow j}(t)$ and $q^{i}(t)$. The derivation, which is close to ours, have been provided in \cite{Altarelli2013}, where an equivalent liner threshold model has been investigated. This model has been also studied in a different setting in a form of the voter model in \cite{Kanoria2011}.

\subsection{Generalized SI model}

Let us now consider the most general case of a unidirectional dynamic model with two states and pairwise interactions between nodes; each of these independent interactions may lead to a transition to the final state. The definition of the generalized SI model in discrete time can be represented as follows:
\begin{align}
& S(i)+S(j)\xrightarrow{\epsilon_{ji}}I(i)+S(j),\\
& S(i)+I(j)\xrightarrow{\lambda_{ji}}I(i)+I(j),\\
& S(i)\xrightarrow{\nu_{i}}I(i).
\end{align}
This diagram represents the dynamic rules at each time step. Here, $i$ and $j$ mean two neighboring nodes in the network, and $\epsilon_{ji}$, $\lambda_{ji}$ and $\nu_{i}$ correspond to transition probabilities at each time step.

Again, since there are only two possible states and the dynamics is unidirectional, the time trajectory of a node $i$ $\left\vert S_{_{0}}SSSSSSI_{_{\tau_{i}}}IIIIII_{_{T}}\right\rangle$ can be parametrized by a single time $\tau_{i}$, when the spin flips from the state $S$ to the state $I$ ($\tau_{i}$ is the first time for which $\sigma^{\tau_{i}}_{i}=I$). If the node $i$ is initially in the state $I$, we set $\tau_{i}=0$, and we put by definition $\tau_{i}=T$ if the flipping happens after the observation time $T$, or never happens.

Therefore, the dynamic cavity equation (\ref{eq:DBP_conditional}) takes the following form for the generalized SI model:

\begin{equation}
m^{i\rightarrow j}(\tau_{i}\mid\tau_{j})=
\hspace{-0.12cm}
\sum_{\{\tau_{k}\}_{k\in\partial i\backslash j}}
\hspace{-0.24cm}
W_{SI}
\hspace{-0.15cm}
\prod_{k\in\partial i\backslash j}
m^{k\rightarrow i}(\tau_{k}\mid\tau_{i}),
\label{eq:message_passing_eq_SI}
\end{equation}
where $W_{SI}$ is the kernel that resumes the dynamics of the model up to the final time $T$, and $\tau_{i} < T$. See Appendix~\ref{app:GSIdetails} for details.

The beliefs $m^{i}(\tau_{i})$, which can be obtained from \eqref{eq:message_passing_eq_SI} in the same way as \eqref{eq:DBP_marginal_conditional} is obtained from \eqref{eq:DBP_conditional}, allow one to define the marginal probabilities describing the dynamics of the SI model:
\begin{align}
& P_{S}^{i}(t)=\sum_{\tau_{i}>t} m^{i}(\tau_{i}),
\\
& P_{I}^{i}(t)=1-P_{S}^{i}(t).
\end{align}
It is also useful to define the marginal probability that node $i$ is in the state $S$ at a given time in the cavity graph $D_{j}$, in which the node $j$ is fixed to the state $S$ for all times:
\begin{equation}
P_{S}^{i \rightarrow j}(t)=\sum_{\tau_{i}>t} m^{i\rightarrow j}(\tau_{i}\mid T).
\label{eq:SI_S_def}
\end{equation}

After some algebra (see Appendix~\ref{app:GSIdetails} for details of derivation), the DMP equations for the generalized SI model take the following form:

\begin{align}
P_{S}^{k \rightarrow i}(t)=P_{S}^{k}(0)(1-\epsilon_{ik})^{t} (1 & - \nu_{k})^{t}\prod_{l\in \partial k \backslash i}\theta^{l \rightarrow k}(t), \label{eq:SIequations:S}
\\
\notag
\theta^{k \rightarrow i}(t)=\theta^{k \rightarrow i}(t-1) - \epsilon_{ki} & \phi_{1}^{k \rightarrow i}(t-1)
\\
& -\lambda_{ki}\phi_{2}^{k \rightarrow i}(t-1),
\\
\notag
\phi_{1}^{k \rightarrow i}(t)=(1-\epsilon_{ki})\phi_{1}^{k \rightarrow i}(t-&1)
\\
-(1-\epsilon_{ki})^{t} (P_{S}^{k \rightarrow i} & (t-1)-P_{S}^{k \rightarrow i}(t)),
\\
\notag
\phi_{2}^{k \rightarrow i}(t)=(1-\lambda_{ki})\phi_{2}^{k \rightarrow i}(t-&1)
\\
+(1-\epsilon_{ki})^{t}(P_{S}^{k \rightarrow i} & (t-1)-P_{S}^{k \rightarrow i}(t)).
\label{eq:SIequations:phi2}
\end{align}

The initial conditions are
\begin{align}
&\theta^{i \rightarrow j}(0)=1,
\\
&\phi_{1}^{i \rightarrow j}(0)=\delta_{\sigma_{i}^{0},S}=P_{S}^{i}(0),
\\
&\phi_{2}^{i \rightarrow j}(0)=\delta_{\sigma_{i}^{0},I}=P_{I}^{i}(0)=1-P_{S}^{i}(0).
\end{align}

As follows from their mathematical definitions, the introduced dynamic variables can be given the following physical interpretations (that follow from their explicit mathematical form):
\begin{itemize}
\item $\theta^{k \rightarrow i}(t)$ is the probability that neither of both $\epsilon$ and $\lambda$ infection signals has been passed from node $k$ to node $i$ up to time $t$ in the cavity dynamics $D_{i}$;
\item $\phi_{1}^{k \rightarrow i}(t)$ is the probability that the $\epsilon$ infection signal has not been passed from node $k$ to node $i$ up to time $t$ in the cavity dynamics $D_{i}$ and that $k$ is in the state $S$ at time $t$;
\item $\phi_{2}^{k \rightarrow i}(t)$ is the probability that  neither of both $\epsilon$ and $\lambda$ infection signals has been passed from node $k$ to node $i$ up to time $t$ in the cavity dynamics $D_{i}$ and that $k$ is in the state $I$ at time $t$;
\item $P_{S}^{k \rightarrow i}(t)$ is the probability that $k$ is in the state $S$ at time $t$ in the cavity dynamics $D_{i}$.
\end{itemize}

\begin{figure}[ht]
\centering
\includegraphics[width=0.84\columnwidth]{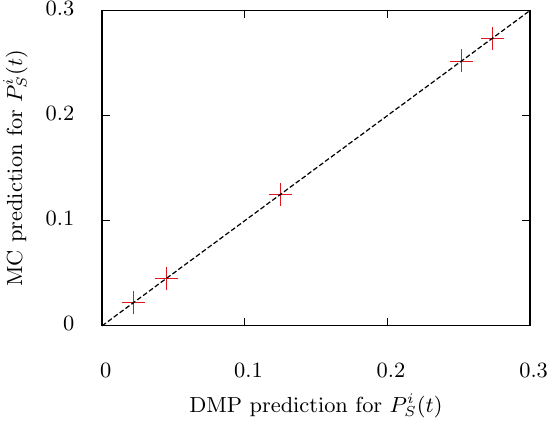}
\caption{(Color online) Comparison of prediction of the DMP equations for the generalized SI model with the Monte Carlo simulations. Marginal probabilities $P_S^{i}(t)$ are presented for five nodes from a tree graph with $N=20$ nodes and $t=5$, the parameters of the model are $\lambda=0.5$, $\epsilon=0.1$ and $\nu=0.1$, there is one infected node at initial time. The MC average is performed over $10^7$ instances. The error bars are negligible and are not shown.}
\label{fig:3}
\end{figure}

Finally, the marginal probabilities for nodes to be in states $S$ or $I$ at time $t$ are computed via
\begin{align}
&
P_{S}^{i}(t)=P_{S}^{i}(0)(1-\nu_{i})^{t}\prod_{k\in \partial i}\theta^{k \rightarrow i}(t),
\label{SI_marginal_factorization}
\\
&
P_{I}^{i}(t)=1-P_{S}^{i}(t).
\end{align}
The exactness of DMP equations for generalized SI model on tree graphs is illustrated in the \fig{fig:3}. Their computational complexity is $\mathcal{O}(N c t)$, where $c$ is the average degree of the graph. Note that the complexity is linear in average degree of the network; this is due to the pairwise nature of the interactions, leading to a factorization of dynamic variables in \eqref{eq:SIequations:S} and \eqref{SI_marginal_factorization}. The same property will also hold for other pairwise models, considered below. These dynamic equations have never appeared in the literature. However, in some sense this model is a straightforward generalization of the SIR model that is considered further, see the next section.

\section{Unidirectional models with $K=3$ states}
\label{sec:K3}

\subsection{SIR model}

The dynamics of the SIR model is defined in discrete time by infection and recovery probabilities, $\lambda_{ij}$ and $\mu_{i}$. At each time step, the following dynamics rules apply \cite{BoccalettiaLatora06}:
\begin{align}
& S(i)+I(j)\xrightarrow{\lambda_{ji}}I(i)+I(j),\label{eq:SIR_rules_StoI}
\\
& I(i)\xrightarrow{\mu_{i}}R(i).
\label{eq:SIR_rules_ItoR}
\end{align}
Note that the SIR model, defined by equations \eqref{eq:SIR_rules_StoI} and \eqref{eq:SIR_rules_ItoR}, represents in some sense a particular case of the generalized SI model, with $\epsilon_{ij}=0$ and $\nu_{i}=0$ for all $i$ and $j$. At the same time, a trivial (independent on the state of neighbors) transition to the $R$ state with probability $\mu_{i}$ is added. Now the time trajectory for a node $i$ can be fully parametrized by two flipping times: $\vec{\sigma}_{i}=\left\vert S_{_{0}}SSSSSSI_{_{\tau_{i}}}IIIIIIR_{_{\omega_{i}}}RRRRRR_{_{T}}\right\rangle \longleftrightarrow (\tau_{i},\omega_{i})$. This leads to the following equations, already derived in a different way (with a correct choice of dynamic variables that have to be used) in \cite{PatientZero2013}:
\begin{align}
P_S^{i \rightarrow j}(t+1)=P_S^{i}(0)\prod_{k\in\partial i \backslash j}&\theta^{k \rightarrow i}(t+1), \label{eq:SIRequations:P_S}
\\
\theta^{k \rightarrow i}(t+1)-\theta^{k \rightarrow i}(t) =
-&\lambda_{ki}\phi^{k \rightarrow i}(t),
\label{eq:SIRequations:theta}
\\
\notag 
\phi^{k \rightarrow i}(t)=(1-\lambda_{ki})(1-\mu_{k}&)\phi^{k \rightarrow i}(t-1)
\\
 -[P_S&^{k \rightarrow i}(t)-P_S^{k \rightarrow i}(t-1)].
\label{eq:SIRequations:phi}
\end{align}
The initial conditions are given by $\theta^{k\rightarrow i}(0)=1$, and $\phi^{k\rightarrow i}(0)=\delta_{\sigma_{k}^{0},I}$. Similarly to the SI model, a concrete physical sense may be given to the dynamic variables $P_S^{k \rightarrow i}(t)$, $\theta^{k \rightarrow i}(t)$ and $\phi^{k \rightarrow i}(t)$ (see \cite{PatientZero2013} for details). The marginal probabilities that node $i$ is in a given state at time $t$ are then given as
\begin{align}
& P_S^{i }(t+1)=P_S^{i}(0)\prod_{k\in\partial i}\theta^{k \rightarrow i}(t+1)\, ,\label{eq:SIRequations:S}
\\
& P_R^{i}(t+1)=P_R^{i}(t)+\mu_{i}P_{I}^{i}(t)\, ,\label{eq:SIRequations:R}
\\
& P_I^{i}(t+1)=1-P_S^{i}(t+1)-P_R^{i}(t+1)\, .\label{eq:SIRequations:I}
\end{align}
The computational complexity of DMP equations for SIR model is $\mathcal{O}(N c t)$. The details of the derivation of DMP equations for the SIR model using dynamic belief propagation are given in the Appendix~\ref{app:SIRdetails}. A numerical study of these equations has been provided in \cite{PatientZero2013} for different type of networks.

Equations reminiscent of (\ref{eq:SIRequations:S}-\ref{eq:SIRequations:I}) were first derived in \cite{KarrerNewman10b} for a more general SIR model with non-exponential transmission and recovery distributions. For this more general case, no easily tractable Markovian form of the DMP is known: the equations in \cite{KarrerNewman10b} are presented in a convolutional form that is complicated for numerical resolution. For a simpler case of constant recovery and transmission rates, the equations of \cite{KarrerNewman10b} simplify. For an ensemble of random graphs with a given probability distribution it is possible to write the mean-field equations on the fraction of nodes in the states $S$, $I$ and $R$ \cite{Volz08,Miller11,lucas2012exact}. These equations are exact in the ensemble of diluted random graphs in the thermodynamic limit, $N\to \infty$ and differ markedly from the naive mean field equations \cite{BoccalettiaLatora06} that completely ignore the topology of the network and therefore provide only a crude approximation to the dynamics. The recent work \cite{shrestha2014message} presented a generalization of the SIR model to the threshold models where a transition happens only if the information is received from a certain number $A$ of neighbors. This model can also be readily solved within the DMP approach. Indeed, the expression \eqref{eq:SIRequations:S} for the marginal probability that node $i$ is in the state $S$ would take a form similar to the second term in the right-hand side of the equation \eqref{eq:RFIM_marginal} in the RFIM, with $q^{k\rightarrow i}(t)$ replaced by $\theta^{k \rightarrow i}(t+1)$: one would need to sum over all the subsets of $\partial i$ that correspond to the transmission of information by at least $A$ neighbors. Hence, this model represents a three-state model with a RFIM-like non-trivial transition to the infected state.

\subsection{Rumor spreading model}

The definition of the rumor spreading model can be summarized as follows \cite{BoccalettiaLatora06}. For the sake of simplicity, we keep the same notations for the states as those used in the SIR model. Each node $i \in V$ at discrete time $t$ can be in one of three states $\sigma^{t}_{i}$: ignorant, $\sigma^{t}_{i}=S$, spreader, $\sigma^{t}_{i}=I$, or stifler, $\sigma^{t}_{i}=R$. At each time step, an ``infected'' node $i$ will recover with probability $1-\prod_{k\in\partial i}(1-\alpha_{ki}\delta_{\sigma^{t}_{k},I})$, and a ``susceptible'' node $i$ will become infected with probability $1-\prod_{k\in\partial i}(1-\lambda_{ki}\delta_{\sigma^{t}_{k},I})$, where $\partial i$ is the set of neighbors of node $i$. The recovered nodes never change their state. These rules can be summarized by the following scheme:
\begin{align}
& S(i)+I(j)\xrightarrow{\lambda_{ji}}I(i)+I(j),\label{eq:RS_rules_StoI}
\\
& I(i)+I(j)\xrightarrow{\alpha_{ji}}R(i)+I(j).
\label{eq:RS_rules_ItoR}
\end{align}
The interpretation of this model is as follows: a spreader node can either inform one of its ignorant neighbors on the rumor, in which case they start to communicate the rumor to their neighbors, or become uninterested in the rumor and turn to the $R$ state if the rumor looses its ``news value''. This happens in a directed way when the spreader gets in contact with another spreader. Note that some rumor spreading models include an additional modeling of such a spreading decay, described by a contact of a spreader with a stifler with the same probability $\alpha$. This additional transition can be easily included in our approach, but for the purpose of this paper we stick to this ``minimal'' version of the ISS model that captures the main features of the rumor spreading process and its difference with respect to epidemiological spreading models, such as the SIR model considered before.

As in the previous cases, the irreversibility of dynamics of the rumor spreading model makes it possible to parametrize the time trajectory of a node $i$ by only two flipping times, $\vec{\sigma}_{i}=(\tau_{i},\omega_{i})$: $\tau_{i}$, indicating a transition from $S$ to $I$ (the first time to be in the state $I$) and $\omega_{i}$, corresponding to a $I$ to $R$ transition (the first time in the state $R$). If the node $i$ is initially in the state $I$, we set $\tau_{i}=0$, and we put by definition $\tau_{i}=T$ if the flipping happens after the termination time $T$, or never happens.

The rumor spreading model, defined via the transition rules (\ref{eq:RS_rules_StoI}-\ref{eq:RS_rules_ItoR}), is notably more complicated than the SIR model because it has two non-trivial transitions, dependent on the state of neighbors. As we will see, it is not easy to obtain the corresponding DMP equations for this model by guessing the correct dynamic variables since the computation of the very DBP messages is required. On the other hand, they appear automatically in the dynamic cavity approach. Let us first state the DMP computational scheme for this model.

The marginal probabilities $P_S^{i}(t+1)$, $P_I^{i}(t+1)$ and $P_R^{i}(t+1)$ that node $i$ is in a state $S$, $I$ and $R$ respectively at time $t$ are given by the following equations that can be iterated in time starting from initial conditions at time $t=0$:
\begin{align}
& P_S^{i}(t+1)=P_S^{i}(0)\prod_{k\in\partial i}\theta^{k \rightarrow i}(t+1)\, ,\label{eq:RSequations:S}
\\
& P_R^{i}(t+1)=P_R^{i}(t)+\sum_{\tau_{i} \leq t}m^{i}(\tau_{i},t+1)\, ,\label{eq:RSequations:R}
\\
& P_I^{i}(t+1)=1-P_S^{i}(t+1)-P_R^{i}(t+1)\, ,\label{eq:RSequations:I}
\end{align}
where $m^{i}(\tau_{i},t+1)$ has a physical meaning of the marginal probability that the node $i$ has switched to the state $I$ at time $\tau_{i}$ and to the state $R$ at time $t+1$. The remaining computational scheme serves to compute this probabilities explicitly. To this purpose, we introduce a number of auxiliary dynamic messages that can be computed iteratively. Again, these messages may be given a physical interpretation, and are defined in the corresponding cavity dynamics. As an illustration, consider the message $P_S^{i\rightarrow j}(t)$, defined as the probability for node $i$ to be in the state $S$ at time $t$ in the cavity graph, in which all the connections of the node $j$, except to $i$, has been removed. It is updated as follows:

\begin{align}
&P_S^{i \rightarrow j}(t+1)=P_S^{i}(0)\prod_{k\in\partial i \backslash j}\theta^{k \rightarrow i}(t+1), \label{eq:RSequations:P_S}
\\
&\theta^{k \rightarrow i}(t+1)-\theta^{k \rightarrow i}(t) =
-\lambda_{ki}\phi^{k \rightarrow i}(t),
\label{eq:RSequations:theta}
\\
\notag
&\phi^{k \rightarrow i}(t)=(1-\lambda_{ki})\phi^{k \rightarrow i}(t-1)+P_S^{k \rightarrow i}(t-1)
\\
&-P_S^{k \rightarrow i}(t)-\sum_{\tau_{k}\leq t-1}(1-\lambda_{ki})^{t-\tau_{k}}m^{k \rightarrow i}(\tau_{k},t\mid T,T).
\label{eq:RSequations:phi}
\end{align}

In these equations, the dynamic messages $\theta^{k \rightarrow i}(t+1)$, $\phi^{k \rightarrow i}(t)$ and $m^{k \rightarrow i}(\tau_{k},t\mid T,T)$ have the following physical sense (for precise mathematical expressions, see Appendix \ref{app:Rumordetails}):
\begin{itemize}
\item $\theta^{k \rightarrow i}(t+1)$ is the probability that the infection signal $\lambda$ has not been passed from node $k$ to node $i$ up to time $t+1$ in the cavity dynamics $D_{i}$;
\item $\phi^{k \rightarrow i}(t)$ is the probability that the infection signal $\lambda$ has not been passed from the node $k$ to the node $i$ up to time $t$ in the cavity dynamics $D_{i}$ and that $k$ is in the state $I$ at time $t$;
\item $m^{k \rightarrow i}(\tau_{k},t\mid T,T)$ is the marginal probability that node $k$ has the trajectory $(\tau_{k},t)$ in the cavity dynamics $D_{i}$.
\end{itemize}
Hence, the last term in the equation \eqref{eq:RSequations:phi} represents a contribution to the change of $\phi^{k \rightarrow i}(t)$ due to the recovery of the node $k$ exactly at time $t$ in the cavity dynamics $D_{i}$. The initial conditions are given by $\theta^{k\rightarrow i}(0)=1$, and $\phi^{k\rightarrow i}(0)=\delta_{\sigma_{k}^{0},I}$. 

So far, the equations (\ref{eq:RSequations:P_S}-\ref{eq:RSequations:phi}) are not in a closed form, we still need to know how to compute $m^{i \rightarrow j}(\tau_{i},t\mid T,T)$ for $\tau_{i}<t$.
We have for each $t$
\begin{align}
\notag
&m^{i \rightarrow j}(0,t\mid \tau_{j},T)=P^{i}_{I}(0)
\\
&\times\Big[ f^{i \rightarrow j}_{\rho_{1},\chi_{1}}(-2,t-2\mid \tau_{j})-f^{i \rightarrow j}_{\rho_{1},\chi_{1}}(-2,t-1\mid \tau_{j}) \Big],
\label{eq:RSequations:m0}
\\
\notag
&m^{i \rightarrow j}(\tau_{i},t\mid \tau_{j},T)=P^{i}_{S}(0)
\\
\notag
&\times\Big[ f^{i \rightarrow j}_{\rho_{1},\chi_{1}}(\tau_{i}-2,t-2\mid \tau_{j})-f^{i \rightarrow j}_{\rho_{1},\chi_{1}}(\tau_{i}-2,t-1\mid \tau_{j})
\\
&-\hspace{-0.05cm}f^{i \rightarrow j}_{\rho_{2},\chi_{2}}(\tau_{i}-\hspace{-0.05cm}1,t-\hspace{-0.05cm}2\hspace{-0.05cm}\mid\hspace{-0.05cm}\tau_{j})\hspace{-0.05cm}
+\hspace{-0.05cm}f^{i \rightarrow j}_{\rho_{2},\chi_{2}}(\tau_{i}-\hspace{-0.05cm}1,t-\hspace{-0.05cm}1\hspace{-0.05cm}\mid\hspace{-0.05cm}\tau_{j}) \Big],
\label{eq:RSequations:m}
\end{align}
for $1 \leq \tau_{i} \leq t-1$ and $0 \leq \tau_{j} \leq t$. The functional $f^{i \rightarrow j}_{\rho,\chi}(t_{1},t_{2}\mid\tau_{j})$ is defined as follows:
\begin{equation}
f^{i \rightarrow j}_{\rho,\chi}(t_{1},t_{2}\hspace{-0.05cm}\mid\hspace{-0.05cm}\tau_{j})=\rho^{j \rightarrow i}(t_{1},t_{2}\hspace{-0.05cm}\mid\hspace{-0.05cm}\tau_{j})\hspace{-0.1cm}\prod_{k\in\partial i \backslash j}\hspace{-0.1cm}\chi^{k \rightarrow i}(t_{1},t_{2}),
\label{eq:RSequations:f}
\end{equation}
where the $\tau_{j}$-dependent coefficients, characterizing the influence of node $j$ on the dynamics of $i$, read for $t_{2}=t-2$ or $t_{2}=t-1$: 
\begin{align}
\notag
&\rho_{1}^{j \rightarrow i}(\tau_{i}-2,t_{2}\mid\tau_{j})
\\
&=(1-\lambda_{ji})^{\tau_{i}-\tau_{j}-1}
\left(\prod_{t'=\tau_{i}}^{t_{2}}\hspace{-0.1cm}\left(1-\alpha_{ji}\mathds{1}[\tau_{j}\leq t']\right)\right),
\label{eq:RSequations:rho1}
\\
\notag
&\rho_{2}^{j \rightarrow i}(\tau_{i}-1,t_{2}\mid\tau_{j})
\\
&=(1-\lambda_{ji})^{\tau_{i}-\tau_{j}}
\left(\prod_{t'=\tau_{i}}^{t_{2}}\hspace{-0.1cm}\left(1-\alpha_{ji}\mathds{1}[\tau_{j}\leq t']\right)\right).
\label{eq:RSequations:rho2}
\end{align}
Let us remind at this point, that the convention \eqref{eq:convention} is used here and in the following. Note that in the update equation \eqref{eq:RSequations:phi} we are only interested in the messages of the form $m^{i \rightarrow j}(\tau_{i},t\mid T,T)$, that correspond to $\tau_{j}=t$, and for which the $j$-influence is not present: in this case $\rho_{1}^{j \rightarrow i}(\tau_{i}-2,t_{2}\hspace{-0.05cm}\mid\hspace{-0.05cm}\tau_{j})$ and $\rho_{2}^{j \rightarrow i}(\tau_{i}-1,t_{2}\hspace{-0.05cm}\mid\hspace{-0.05cm}\tau_{j})$ are simply equal to one. Still, in the computation scheme for $\chi_{1}^{k \rightarrow i}(\tau_{i}-2,t_{1})$ and $\chi_{2}^{k \rightarrow i}(\tau_{i}-1,t_{1})$ all the values $0 \leq \tau_{j} \leq t-1$ are also required, since the remaining update equations read
\begin{align}
\notag
\chi_{1}^{k \rightarrow i}(\tau_{i}-2,t-1) = \chi_{1}^{k \rightarrow i}&(\tau_{i}-2,t-2) 
\\
-&\alpha_{ki}\psi_{1}^{k \rightarrow i}(\tau_{i}-2,t-1),
\label{eq:RSequations:chi1}
\\
\notag
\chi_{2}^{k \rightarrow i}(\tau_{i}-1,t-1) = \chi_{2}^{k \rightarrow i}&(\tau_{i}-1,t-2)
\\ 
-&\alpha_{ki}\psi_{2}^{k \rightarrow i}(\tau_{i}-1,t-1),
\label{eq:RSequations:chi2}
\end{align}
and
\begin{align}
\notag
\psi_{1}^{k \rightarrow i}(\tau_{i}-2,t-1) = P&_S^{k \rightarrow i}(t-2 \mid \tau_{i})
\\
\notag
-P_S^{k \rightarrow i}(t-1 \mid \tau_{i})+(1-&\alpha_{ki}\mathds{1}_{\tau_{i} \neq t-1})\psi_{1}^{k \rightarrow i}(\tau_{i}-2,t-2)
\\
\notag
-\sum_{\tau_{k}\leq t-2}(1-\lambda_{ki})^{\tau_{i}-\tau_{k}-1}
&\left(\prod_{t'=\tau_{i}}^{t-2}(1-\alpha_{ki}\mathds{1}[\tau_{k}\leq t'])\right)
\\
&\times m^{k \rightarrow i}(\tau_{k},t-1\mid \tau_{i},T),
\label{eq:RSequations:psi1}
\end{align}
\begin{align}
\notag
\psi_{2}^{k \rightarrow i}(\tau_{i}-1,t-1)=P&_S^{k \rightarrow i}(t-2 \mid \tau_{i})
\\
\notag
-P_S^{k \rightarrow i}(t-1 \mid \tau_{i})+(1-&\alpha_{ki}\mathds{1}_{\tau_{i} \neq t-1})\psi_{2}^{k \rightarrow i}(\tau_{i}-1,t-2)
\\
\notag
-\sum_{\tau_{k}\leq t-2}(1-\lambda_{ki})^{\tau_{i}-\tau_{k}}
&\left(\prod_{t'=\tau_{i}}^{t-2}\left(1-\alpha_{ki}\mathds{1}[\tau_{k}\leq t']\right)\right)
\\
&\times m^{k \rightarrow i}(\tau_{k},t-1\mid \tau_{i},T).
\label{eq:RSequations:psi2}
\end{align}
The conditional quantity $P_S^{k \rightarrow i}(t_{1} \mid \tau_{i})$ is defined as
\begin{equation}
P_S^{k \rightarrow i}(t_{1} \mid \tau_{i})=P_S^{k}(0)(1-\lambda_{ik})^{t_{1}-\tau_{i}}\prod_{l\in\partial k \backslash i}\theta^{l \rightarrow k}(t_{1}).
\label{eq:RSequations:PS_conditioned}
\end{equation}
The necessary initial conditions are given by $\chi_{1}^{k \rightarrow i}(-2,-1)=1$ and $\psi_{1}^{k \rightarrow i}(-2,0)=\phi^{k \rightarrow i}(0)$. The following border conditions are used for $\tau_{i}=t-1$:
\begin{align}
&\chi_{1}^{k \rightarrow i}(t-3,t-2)=\theta^{k \rightarrow i}(t-2),
\label{eq:RSequations:border_condition_chi1}
\\
&\chi_{2}^{k \rightarrow i}(t-2,t-2)=\theta^{k \rightarrow i}(t-1),
\label{eq:RSequations:border_condition_chi2}
\\
&\psi_{1}^{k \rightarrow i}(t-3,t-2)=\phi^{k \rightarrow i}(t-2),
\label{eq:RSequations:border_condition_psi1}
\\
&\psi_{2}^{k \rightarrow i}(t-2,t-2)=(1-\lambda_{ki})\phi^{k \rightarrow i}(t-2).
\label{eq:RSequations:border_condition_psi2}
\end{align}
and $\chi_{1}^{k \rightarrow i}(t-3,t-1)$, $\chi_{2}^{k \rightarrow i}(t-2,t-1)$, $\psi_{1}^{k \rightarrow i}(t-3,t-1)$, $\psi_{2}^{k \rightarrow i}(t-2,t-1)$ follow the equations (\ref{eq:RSequations:chi1}-\ref{eq:RSequations:psi2}). 

Therefore, the computation of $m^{i \rightarrow j}(\tau_{i},t\mid T,T)$ for $\tau_{i}<t$ involves messages $m^{i \rightarrow j}(\tau_{i},t-1\mid \tau_{j},T)$ for $\tau_{i}<t-1$ and $\tau_{j} \leq t-1$, computed at a previous step. Finally, the marginal probabilities $m^{i}(\tau_{i},t+1)$ are computed via equations \eqref{eq:RSequations:m0} and \eqref{eq:RSequations:m}, with replacement of the indices $i \rightarrow j$ simply by $i$, and the corresponding change of product over $k\in\partial i \backslash j$ in the definition \eqref{eq:RSequations:f} by the product over all the neighboring nodes $k\in\partial i$. The computational complexity of DMP equations for the rumor spreading model is $\mathcal{O}(N c t^3)$, where $c$ is the average degree of the graph. The details of the derivation are presented in the Appendix~\ref{app:Rumordetails}.

\begin{figure}[!ht]
\centering
\includegraphics[width=0.97\columnwidth]{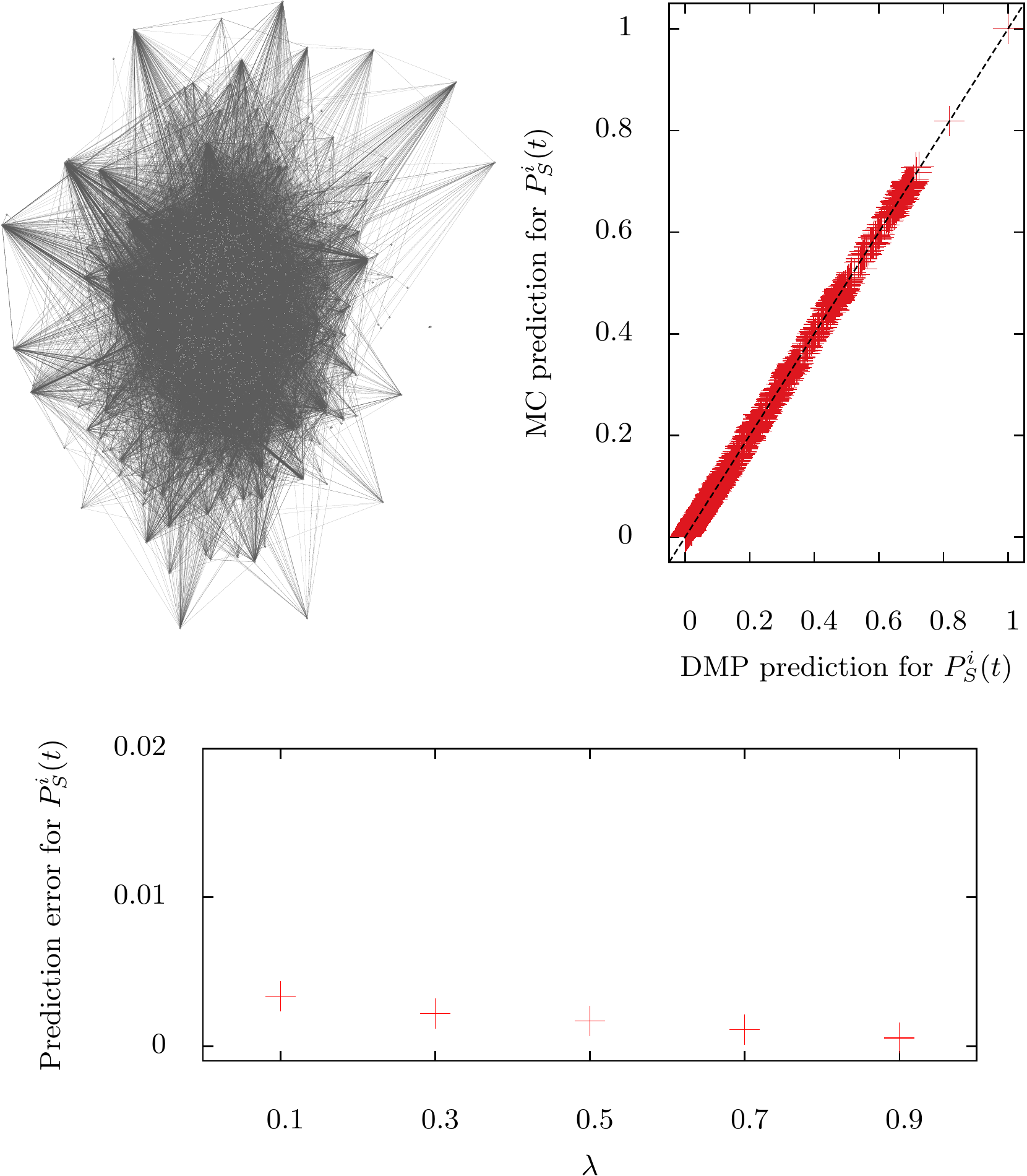}
\caption{(Color online) Top right: Comparison of prediction of the DMP equations for the rumor spreading model with the Monte Carlo (MC) simulations in a typical case. Marginal probabilities $P_S^{i}(t)$ are presented for the Facebook-like social network with $N=1899$ nodes and $t=10$, the parameters of the model are $\lambda=0.3$, $\alpha=0.2$, there is one infected node at initial time. Top left: A representation of the topology of the network, generated with Gephi \cite{bastian2009gephi}. The high-degree nodes (hubs) are placed on the periphery. Bottom: Study of the average prediction error per node for marginal probabilities $P_S^{i}(t)$ as a function of $\lambda$. In this plot, $t=10$ and $\alpha=0.2$, so that the point $\lambda=0.3$ corresponds to the comparison above. For both plots, the MC average is performed over $10^4$ instances. The error bars are smaller than the symbol size on the plots and are not shown.}
\label{fig:4}
\end{figure}

\begin{figure}[!ht]
\centering
\includegraphics[width=0.97\columnwidth]{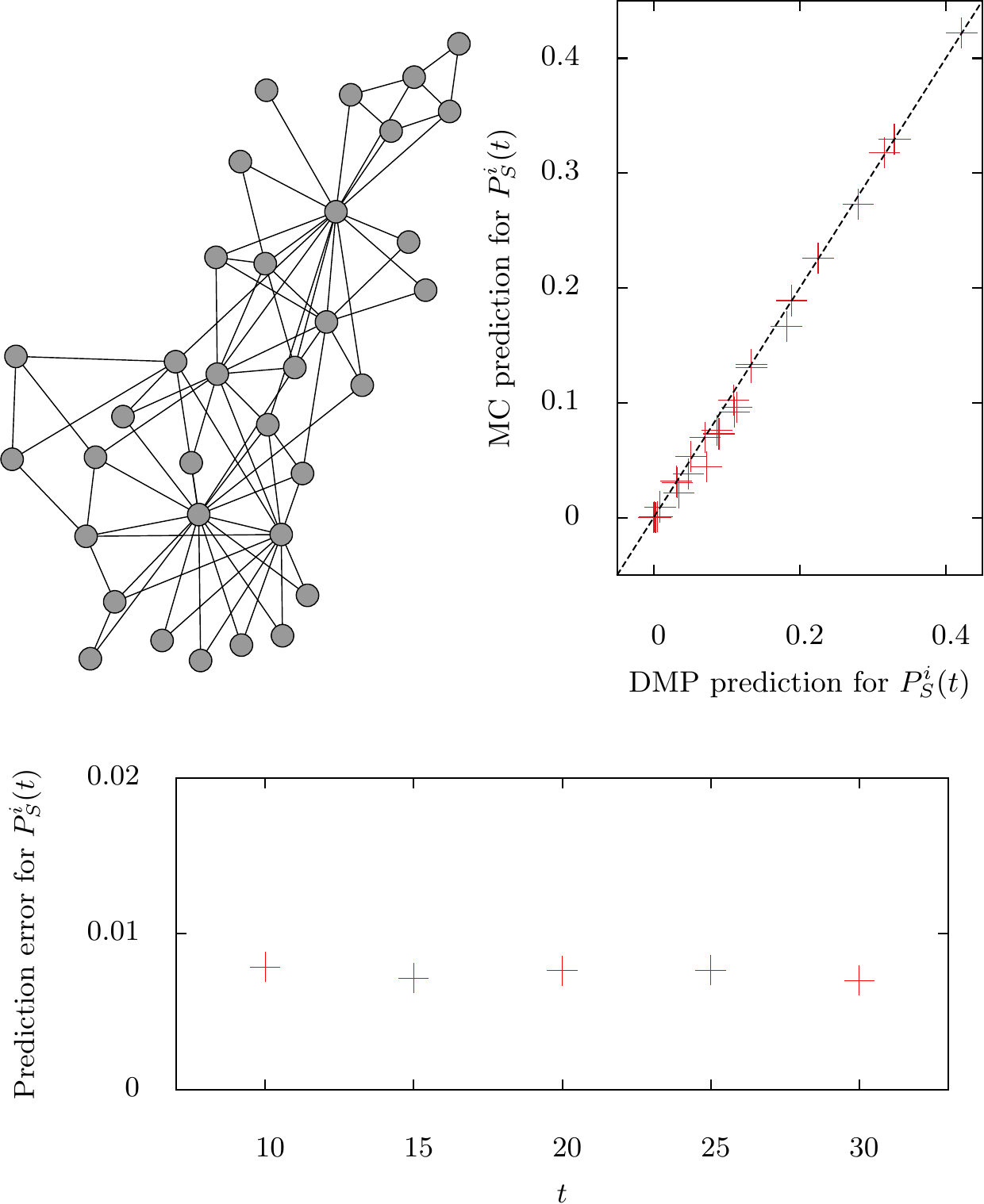}
\caption{(Color online) Top right: Comparison of prediction of the DMP equations for the rumor spreading model with the Monte Carlo (MC) simulations in a typical case. Marginal probabilities $P_S^{i}(t)$ are presented for the Zachary's karate club network with $N=34$ nodes and $t=10$, the parameters of the model are $\lambda=0.3$, $\alpha=0.2$, there is one infected node at initial time. Top left: A representation of the topology of the network, generated with Gephi \cite{bastian2009gephi}. This network has a block structure and contains many loops of small length. Bottom: Study of the average prediction error per node for marginal probabilities $P_S^{i}(t)$ as a function of time $t$. In this plot, $\lambda=0.3$ and $\alpha=0.2$, so that the point $t=10$ corresponds to the comparison above. For both plots, the MC average is performed over $10^4$ instances. The error bars are smaller than the symbol size on the plots and are not shown.}
\label{fig:5}
\end{figure}

The validity of these equations has been checked numerically via comparison with the Monte Carlo (MC) simulation: the marginals given by the DMP equations appear to be exact on any tree graph. Although \emph{a priori} the DMP equations are not guaranteed to be exact on networks that do not have a locally tree-like structure, they provide remarkably accurate predictions even for small and loopy networks. For example, we have tested the performance of the DMP equations for the rumor spreading model on two real-world networks. The first example is a Facebook-like social network with 1899 nodes and 20 296 edges that represents an online community for students at University of California, Irvine \cite{Opsahl2009}; the predictions for the marginals given by DMP are compared with the values obtained from $10^4$ MC simulations, see \fig{fig:4}. Another test has been performed for the small Zachary's karate club network of friendships between 34 members of a karate club at a US university \cite{Zachary77}, the results are presented in the \fig{fig:5}. In both cases, the predictions of the DMP equations appear to be very accurate with respect to the true values of the marginals.

The DMP equations for the rumor spreading model have never been reported so far. The existing approaches include the naive mean-field equations \cite{BoccalettiaLatora06} that are derived under the complete-mixing assumption and completely neglect the topology of the network, or the so-called heterogeneous mean-field equations \cite{vespignani2012modelling} that assume equivalent behavior for different nodes of the same degree; they are averaged over the ensemble of random graphs and are not applicable on a single instance of the network.

\section{Unidirectional models with $K>3$ states}
\label{sec:K4}

The DMP equations for the rumor spreading model, described in the previous section, can be easily generalized to a more complicated pairwise model with three states, similar to the generalized SI model.
In this section we will illustrate the procedure for deriving the DMP equations for pairwise models with a larger number of states, using as an example a ``minimal'' model with $K=4$ states, which is an extension of the rumor spreading model with an additional non-trivial transition to the final state. 
The procedure for deriving these equations from the general dynamic cavity equation \eqref{eq:DBP_conditional} is very similar to the derivation of the DMP equations for the rumor spreading model. A generalization for any larger number of states seems to be straightforward. Note that the models with $K \geq 3$ states that include direct transitions that skip some number of intermediate states can also be taken into account in this approach.

Let us consider four states $S$, $I_{1}$, $I_{2}$ and $R$, and the following dynamic rules:
\begin{align}
& S(i)+I_{1}(j)\xrightarrow{\lambda_{ji}}I_{1}(i)+I_{1}(j),\label{eq:Kmodel_rules_StoI1}
\\
& I_{1}(i)+I_{1}(j)\xrightarrow{\alpha_{ji}}I_{2}(i)+I_{1}(j),\label{eq:Kmodel_rules_I1toI2}
\\
& I_{2}(i)+I_{2}(j)\xrightarrow{\beta_{ji}}R(i)+I_{2}(j).
\label{eq:Kmodel_rules_I2toR}
\end{align}

Now the time trajectory of the node $i$ can be parametrized by three flipping times: $\tau_{i}$ (first time in $I_{1}$), $\omega_{i}$ (first time in $I_{2}$) and $\varepsilon_{i}$ (first time in $R$). The trajectory of the spin $i$ is hence described by $\vec{\sigma}_{i}(t)=(\tau_{i},\omega_{i},\varepsilon_{i})$, and the corresponding marginal of the dynamic cavity equation \eqref{eq:DBP_marginal_conditional} could be written as $m^{i}(\tau_{i},\omega_{i},\varepsilon_{i})$. Similarly to the SI, the SIR and the rumor spreading model, we might expect that the expressions for the marginal probabilities at time $t$ could be written in the following form:
\begin{align}
& P_S^{i }(t+1)=P_S^{i}(0)\prod_{k\in\partial i}\theta^{k \rightarrow i}(t+1)\, ,\label{eq:Kequations:S}
\\
& P_R^{i}(t+1)=P_R^{i}(t)+\sum_{\substack{\omega_{i} \leq t \\ \tau_{i}+1 \leq \omega_{i}}}m^{i}(\tau_{i},\omega_{i},t+1)\, ,\label{eq:Kequations:R}
\\
& P_{I_{2}}^{i}(t+1)=P_{I_{2}}^{i}(t)+\sum_{\substack{\tau_{i} \leq t \\ \varepsilon_{i} > t+1}}m^{i}(\tau_{i},t+1,\varepsilon_{i})\, ,\label{eq:Kequations:I2}
\\
& P_{I_{1}}^{i}(t+1)=1-P_S^{i}(t+1)-P_{I_{2}}^{i}(t+1)-P_R^{i}(t+1)\, .\label{eq:Kequations:I1}
\end{align}
The apparent difficulty in the equation \eqref{eq:Kequations:I2} is that the sum runs over all the flipping times $\varepsilon_{i}>t+1$, and the number of terms can potentially be very big, of the order of the stopping time $T$. In the Appendix \ref{app:K4details} it is shown that this difficulty can be overcome if one defines a new sort of messages:
\begin{equation}
\mu^{k \rightarrow i}(\tau_{k},t \mid T,T) = \sum_{\varepsilon_{k} \geq t+1} m^{k \rightarrow i}(\tau_{k},t,\varepsilon_{k}\mid T,T,T).
\end{equation}
The evolution of $\mu^{k \rightarrow i}(\tau_{k},t \mid T,T)$ follows the same equations as for the rumor spreading model (\ref{eq:RSequations:m0}-\ref{eq:RSequations:border_condition_psi2}), except that now we will require the computation of ${\mu^{k \rightarrow i}(\tau_{k},t\mid\tau_{i},\omega_{i})}$. The details of derivation for this case is presented in the Appendix \ref{app:K4details}.

The generalization of this model may describe different models with 4 states, for example the generalization of the SIR model that include immunized or exposed states \cite{anderson1991infectious}. As it has been expected, the computational complexity for this $K=4$ model is higher: $\mathcal{O}(N c t^5)$, where $c$ is the average degree of the graph. From the structure of the solution for this minimal model, which makes use of the equations for the model with a lower number of steps, we conjecture that a general model with unidirectional dynamics and $M$ non-trivial transitions will have the computational complexity growing as $t^{2M-1}$. Note that $M$ is not always equal to $K-1$, for instance, compare the SIR ($K=3$, $M=1$) and the rumor spreading ($K=3$, $M=2$) models.

\section{Conclusion}

In this paper we have developed a general approach for deriving the dynamic message-passing equations that describe models with unidirectional dynamics and an arbitrary number of states. These equations can be iterated in physical time starting from arbitrary initial conditions, and allow one to estimate the exact values of marginal probabilities on locally tree-like graphs, providing good approximation for real-world networks. These closed-form equations can be derived starting from the dynamic belief propagation equation on time trajectories, and using the causality and the normalization properties of messages. The dynamic variables that appear in the resulting DMP equations are typically represented by the weighted sums of dynamic BP messages, and emerge automatically in this approach.

Importantly, although the general formulation of dynamic BP on trajectories is of exponential complexity, it takes only a polynomial number of steps $t^{2M-1}$ (where $M$ is the number of non-trivial transitions in the considered models) to solve the corresponding DMP equations. The growth of the number of operations with the number of states of the model in the DMP equations is essentially due to the local effects of retro-action that have to be taken into account. It would be interesting to understand whether there is a way to reduce further this computational complexity, or to prove that this is the minimal number of operations required in order to provide exact equations for these models on tree networks.

The DMP approach opens a way to a number of applications aimed at a better control of the cascading processes on networks. Since the transmission probabilities can be time-dependent, the DMP equations can be used for the dynamically changing graphs. The fact that the DMP equations can be applied to a single instance of a graph has been recently used for the algorithmic application to the inverse problem in the context of the epidemic spreading: the inference of the epidemic origin of an epidemic outbreak \cite{PatientZero2013}. The DMP equations are also promising for optimization and control in models that incorporate the changes of individuals' behavior during the dynamical process \cite{BoccalettiaLatora06,vespignani2012modelling}. Polynomial-complexity DMP equations for the forward dynamics could be used on their own in these potential applications. Another possible strategy would consist in adding extra check nodes (corresponding to the optimization constraints) directly in the factor graph, exemplified in the \fig{fig:2}. This may, however, lead to convergence issues of the corresponding dynamic BP equations for sufficiently complex optimization problems; the discussion of this point is beyond the scope of the present work. These and other applications of the DMP approach to optimization problems are left for future work.

\begin{acknowledgments}
The authors are grateful to Hiroki Ohta and Silvio Franz for fruitful discussions and valuable comments. 
This work has been supported by the Grant DySpaN of
Triangle de la Physique.
\end{acknowledgments}



\onecolumngrid
\appendix

\section{Belief propagation equations for static problems}
\label{app:BPstatic}

Let us assume that a static problem is formulated in terms of a graphical model, defined on a tree graph by a joint probability distribution
\begin{equation}
P(\underline{\sigma})=\frac{1}{Z}\prod_{a=1}^{M}\psi_{a}(\underline{\sigma}_{\partial a}),
\label{eq:model_joint_probability}
\end{equation}
where $Z$ is the normalization constant. Note that this expression is given in a factorized form, each factor $\psi_{a}$ representing a local interaction weight. Very often it is convenient to represent the graphical model in a form of a corresponding factor graph that reflects this structure of the model \eqref{eq:model_joint_probability}. The factor graph can be thought of as a bipartite graph $G=(V,F,E)$: $V$ is a list of variables, $\underline{\sigma}=\{\sigma_{i}\}_{i \in V}$, and $F$ represents a list of interactions, or function nodes, so that an edge $(i,a)\in E$ is present if the interaction $a\in F$ involves a variable $\sigma_{i}$ in the node $i\in V$. We assume here that the set of possible values of  $\sigma_{i}$ is a finite set of size $K$ ($K=2$ for binary spins or boolean variables, $K>3$ for Potts spins or colors, \emph{etc}.). The neighboring nodes in the factor graphs are denoted by $\partial i$ and $\partial a$ for the variables and function nodes, correspondingly. The marginal probability distribution (also called belief) that the variable on the site $i$ takes value $\sigma_{i}$, is defined as 
\begin{equation}
m^{i}(\sigma_{i})=\sum_{\underline{\sigma}_{\backslash i}}P(\underline{\sigma}).
\label{eq:BP_marginal_definition}
\end{equation}
The basic idea of the belief propagation method (also known as the sum-product algorithm) consists in the following observation: since the model is defined on a tree, when one removes a site $i$ from the graph $G$ and cuts the corresponding connection to the neighboring interactions, the resulting cavity graph $G^{(i)}$ is given by a collection of independent and statistically uncorrelated branches of a tree. Therefore, the marginal $m^{i}(\sigma_{i})$ can be expressed simply as a product over the conditional probabilities that represent the contributions of these branches:
\begin{equation}
m^{i}(\sigma_{i})=\frac{1}{Z^{i}}\prod_{a \in \partial i}\widehat{m}^{a \rightarrow i}(\sigma_{i}).
\label{eq:BP_marginal}
\end{equation}
In this expression, $Z^{i}$ is the normalization factor that ensures that $\sum_{\sigma_{i}}m^{i}(\sigma_{i})=1$, and $\partial i$ stands for the neighbors of $i$ in the factor graph. The probability $\widehat{m}^{a \rightarrow i}(\sigma_{i})$, called the message, is defined as the marginal probability that node $i$ takes value $\sigma_{i}$ in the modified graph, in which all the interactions around $i$ except $a$ have been cut out. Now, in order to compute $m^{i}(\sigma_{i})$, one needs to know the values of $\widehat{m}^{a \rightarrow i}(\sigma_{i})$. These quantities obey the coupled self-consistency equations \cite{YedidiaFreeman2003,MezardMontanari2009}  
\begin{align}
&m^{i\rightarrow a}(\sigma_{i})=\frac{1}{Z^{i\rightarrow a}}\prod_{b \in \partial i \backslash a}\widehat{m}^{b \rightarrow i}(\sigma_{i}),
\label{eq:BP1}
\\
&\widehat{m}^{a\rightarrow i}(\sigma_{i})=\frac{1}{Z^{a\rightarrow i}}\sum_{\underline{\sigma}_{\partial a \backslash i}}\psi_{a}(\underline{\sigma}_{\partial a})\prod_{k \in \partial a \backslash i}m^{k \rightarrow a}(\sigma_{k}),
\label{eq:BP2}
\end{align}
where we have also introduced another sort of messages, $m^{i\rightarrow a}(\sigma_{i})$, which is defined as the marginal probability that node $i$ takes value $\sigma_{i}$ in the modified graph, in which the interaction $a$ has been deleted. In these equations, $Z^{i\rightarrow a}$ and $Z^{a\rightarrow i}$ are the normalization constants. The coupled equations \eqref{eq:BP1} and \eqref{eq:BP2} are usually solved by iteration: first, one initialises all the messages to some value, and iterate the equations \eqref{eq:BP1} and \eqref{eq:BP2} until convergence. Then, the final values for the messages $\widehat{m}^{a \rightarrow i}(\sigma_{i})$ are used in \eqref{eq:BP_marginal} for computing the exact values of the marginal probability distributions $m^{i}(\sigma_{i})$. This procedure explains the fine terminology of the BP algorithm: the iteration of equations \eqref{eq:BP1} and \eqref{eq:BP2} can be thought of as a message-passing protocol, each message holding a conditional information on the probability of the corresponding variable; the marginal is then given by a belief shaped by the information contained in all the messages arriving to the node.   

Note that although the equations (\ref{eq:BP_marginal})-(\ref{eq:BP2}) have been derived for a tree graph, they can  be viewed as an algorithm that can be run on an arbitrary interaction graph. They will provide accurate estimations of the marginals as long as the replica symmetric assumption holds for the interaction graph, i.e. that the correlations induced by loops decay fast enough, so that the approximation \eqref{eq:BP_marginal} as a product over neighboring interactions is correct (see \cite{MezardMontanari2009} for more details). In particular, the BP equations (\ref{eq:BP_marginal})-(\ref{eq:BP2}) give asymptotically exact (in the thermodynamic limit $N \to \infty$) expressions for the beliefs on the only locally tree-like networks; sparse random graphs, as well as many real-world networks of interest, fall into this category. 

Sometimes, it is easier to eliminate one sort of messages in \eqref{eq:BP1} and \eqref{eq:BP2} and to use a single iteration equation for messages instead of two:
\begin{equation}
\widehat{m}^{a\rightarrow i}(\sigma_{i})=\frac{1}{Z^{\rightarrow i}}\sum_{\underline{\sigma}_{\partial a \backslash i}}\psi_{a}(\underline{\sigma}_{\partial a})\prod_{k \in \partial a \backslash i}\prod_{b \in \partial k \backslash a}\widehat{m}^{b \rightarrow k}(\sigma_{k}),
\label{eq:BP}
\end{equation}
with $Z^{\rightarrow i}=Z^{a\rightarrow i}\prod_{k \in \partial a \backslash i}Z^{k\rightarrow a}$. This expression further simplifies in the important case of pairwise models, when the variables interact pairs by pairs, and the joint probability distribution factorizes over the links in the graph:
\begin{equation}
P(\underline{\sigma})=\frac{1}{Z}\prod_{(ij)}\psi_{ij}(\sigma_{i},\sigma_{j}).
\end{equation}
In this case, the update equation \eqref{eq:BP} can be rewritten exclusively in terms of messages $m^{i\rightarrow j}(\sigma_{i})$, a short-cut for $m^{i\rightarrow (ij)}(\sigma_{i})$: 
\begin{equation}
m^{i\rightarrow j}(\sigma_{i})=\frac{1}{Z^{i\rightarrow j}}\prod_{k \in \partial i \backslash j}\sum_{\sigma_{k}}\psi_{ik}(\sigma_{i},\sigma_{k})m^{k\rightarrow i}(\sigma_{k}).
\label{eq:BP_pairwise}
\end{equation}

\section{Properties of the dynamic belief propagation equations}
\label{app:DBPdetails}

The exact BP equation for the joint probability distribution of pairs of time trajectories (\ref{eq:Joint_proba_distri_pairs}) in terms of messages $m^{i\rightarrow j}(\vec{\sigma}_{i},\vec{\sigma}_{j})$ reads:

\begin{equation}
m^{i\rightarrow j}(\vec{\sigma}_{i},\vec{\sigma}_{j})=\frac{1}{Z^{i\rightarrow j}}\sum_{\{\vec{\sigma}_{k}\}_{k\in\partial i\backslash j}}\left[\prod_{t=0}^{T-1}w_{i}(\sigma_{i}^{t+1}\mid\{\sigma_{k}^{t}\}_{k\in\partial i\backslash j},\sigma_{j}^{t})P(\{\sigma_{i}^{0}\}_{i\in V})\right]\prod_{k\in\partial i\backslash j}m^{k\rightarrow i}(\vec{\sigma}_{k},\vec{\sigma}_{i}).\label{eq:DBP_general}
\end{equation}

The normalization constant $Z^{i\rightarrow j}$ can be calculated explicitly for the Markovian dynamics from the normalization condition
\begin{equation}
\sum_{\vec{\sigma}_{i}, \vec{\sigma}_{j}} m^{i\rightarrow j}(\vec{\sigma}_{i},\vec{\sigma}_{j})=1
\end{equation}
For example, in the case of general Markov dynamics we use the fact that $m^{k\rightarrow i}(\vec{\sigma}_{k},\vec{\sigma}_{i})$ does not depend on $\sigma_{i}^T$ and perform the summation first over $\sigma_{j}^T$, then over $\sigma_{i}^T$, and so on for the times $T-1,\ldots,0$. Finally, we get the normalization factor
\begin{equation}
Z^{i\rightarrow j}=\frac{1}{2^{(T+1)(d_{i}-2)}}
\end{equation}
for this case, where $d_{i}$ is the number of neighbors of the node $i$ in the initial graph.

The message $m^{i\rightarrow j}(\vec{\sigma}_{i},\vec{\sigma}_{j})$ has a meaning of probability for the trajectories $\vec{\sigma}_{i}, \vec{\sigma}_{j}$ in the transformed cavity graph, where the factor node $j$ has been removed. We can also rewrite the equation (\ref{eq:DBP_general}) in terms of conditional probabilities $m^{i\rightarrow j}(\vec{\sigma}_{i}\mid\vec{\sigma}_{j})$ on the cavity graph. Thus, for the dynamics obeying the Markov property we get
\begin{equation}
\sum_{\vec{\sigma}_{i}} m^{i\rightarrow j}(\vec{\sigma}_{i},\vec{\sigma}_{j})=\frac{1}{2^{T+1}},
\end{equation}
and hence recover the equations \eqref{eq:DBP_conditional} and \eqref{eq:DBP_marginal_conditional}:

\begin{equation}
m^{i\rightarrow j}(\vec{\sigma}_{i}\mid\vec{\sigma}_{j})=\sum_{\{\vec{\sigma}_{k}\}_{k\in\partial i\backslash j}}\left[\prod_{t=0}^{T-1}w_{i}(\sigma_{i}^{t+1}\mid\{\sigma_{k}^{t}\}_{k\in\partial i\backslash j},\sigma_{j}^{t})P(\{\sigma_{i}^{0}\}_{i\in V})\right]\prod_{k\in\partial i\backslash j}m^{k\rightarrow i}(\vec{\sigma}_{k}\mid\vec{\sigma}_{i}),
\label{eq:DBP_conditional_app}
\end{equation}
\begin{equation}
m^{i}(\vec{\sigma}_{i})=\sum_{\{\vec{\sigma}_{k}\}_{k\in\partial i}}\left[\prod_{t=0}^{T-1}w_{i}(\sigma_{i}^{t+1}\mid\{\sigma_{k}^{t}\}_{k\in\partial i})P(\{\sigma_{i}^{0}\}_{i\in V})\right]\prod_{k\in\partial i}m^{k\rightarrow i}(\vec{\sigma}_{k}\mid\vec{\sigma}_{i}).
\label{eq:DBP_marginal_conditional_app}
\end{equation}

The message $m^{i\rightarrow j}(\vec{\sigma}_{i}\mid\vec{\sigma}_{j})$ has a meaning of probability for the trajectory $\vec{\sigma}_{i}$ given the trajectory $\vec{\sigma}_{j}$ in the transformed cavity graph, where the factor node $j$ has been removed. The normalization factor in this equation is exactly equal to $1$, due to the Markov property of the dynamics. Note that, again, by construction, $m^{k\rightarrow i}(\vec{\sigma}_{k}\mid\vec{\sigma}_{i})$ does not depend on $\sigma_{i}^T$, so this variable can be erased. Then, as far as $\prod_{t=1}^{T-1}w_{i}(\sigma_{i}^{t+1}\mid\{\sigma_{k}^{t}\}_{k\in\partial i\backslash j},\sigma_{j}^{t})$ does not depend on $\sigma_{k}^T$, we can perform the sum over $\sigma_{k}^T$ in the right-hand side of the equation (\ref{eq:DBP_conditional_app}), which gives exactly the Kanoria-Montanari expression \cite{Kanoria2011}:

\begin{equation}
m^{i\rightarrow j}_{T+1}(\vec{\sigma}_{i}\mid\vec{\sigma}_{j})=\sum_{\{\sigma_{k}^{0},\ldots,\sigma_{k}^{T-1}\}_{k\in\partial i\backslash j}}\left[\prod_{t=0}^{T-1}w_{i}(\sigma_{i}^{t+1}\mid\{\sigma_{k}^{t}\}_{k\in\partial i\backslash j},\sigma_{j}^{t})P(\{\sigma_{i}^{0}\}_{i\in V})\right]\prod_{k\in\partial i\backslash j}m^{k\rightarrow i}_{T}(\vec{\sigma}_{k}\mid\vec{\sigma}_{i}),\label{eq:Kanoria_Montanari}
\end{equation}
where we denote $m^{i\rightarrow j}_{T+1}(\vec{\sigma}_{i}\mid\vec{\sigma}_{j})=m^{i\rightarrow j}(\vec{\sigma}_{i}\mid\vec{\sigma}_{j})$ and $m^{i\rightarrow j}_{T}(\vec{\sigma}_{i}\mid\vec{\sigma}_{j})=\sum_{\sigma_i^{T+1}}m^{i\rightarrow j}_{T+1}(\vec{\sigma}_{i}\mid\vec{\sigma}_{j}).$

\subsection*{Purely directed case}

In the purely directed case, the direct influence between neighboring nodes $i$ and $j$ runs only in one direction. It means that $m^{i\rightarrow j}(\vec{\sigma}_{i}\mid\vec{\sigma}_{j})$ does not depend on the variable $\vec{\sigma}_{j}$, and the equation \eqref{eq:DBP_conditional_app} is reduced to

\begin{equation}
m^{i\rightarrow j}(\vec{\sigma}_{i})=\sum_{\{\vec{\sigma}_{k}\}_{k\in\partial_{in} i}}\left[\prod_{t=0}^{T-1}w_{i}(\sigma_{i}^{t+1}\mid\{\sigma_{k}^{t}\}_{k\in\partial_{in} i})P(\{\sigma_{i}^{0}\}_{i\in V})\right]\prod_{k\in\partial_{in} i}m^{k\rightarrow i}(\vec{\sigma}_{k}),
\label{eq:DBP_directed}
\end{equation}
where $\partial_{in} i$ denotes the set of neighbors of $i$ that have a direct influence on $i$.
Therefore, writing the marginal in a factorized form
\begin{equation}
m^{i}(\vec{\sigma}_{i})=\prod_{t}m^{i}_{t}(\sigma_{i}^{t}),
\label{eq:DBP_time_factorization}
\end{equation}
we immediately get from \eqref{eq:DBP_conditional_app} for $t>0$
\begin{equation}
m^{i}_{t+1}(\sigma_{i}^{t+1})= \sum_{\{\sigma_{k}^{t}\}_{k \in \partial_{in} i}} w_{i}(\sigma_{i}^{t+1}\mid\{\sigma_{k}^{t}\}_{k\in\partial_{in} i})\prod_{k\in\partial_{in} i}m^{k}_{t}(\sigma_{k}^{t}).
\end{equation}
Note, however, that in the case of the dynamics in which the state of a node depends on the state of the same node at previous time, the factorization \eqref{eq:DBP_time_factorization} does not lead to a decoupled expression, and we have
\begin{equation}
\prod_{t}m^{i}_{t+1}(\sigma_{i}^{t+1})=\prod_{t}\left[\sum_{\{\sigma_{k}^{t}\}_{k \in \partial_{in} i}} w_{i}(\sigma_{i}^{t+1}\mid\{\sigma_{k}^{t}\}_{k\in\partial_{in} i},\sigma_{i}^{t})P(\{\sigma_{i}^{0}\}_{i\in V})\prod_{k\in\partial_{in} i}m^{k}_{t}(\sigma_{k}^{t})\right].
\end{equation}

\section{Derivation of the DMP equations for the zero-temperature RFIM}
\label{app:RFIMdetails}

Using the dynamic properties of the model, the equation (\ref{eq:DBP_conditional}) for $\tau_{i} < T$ can be expressed as follows:
\begin{equation}
m^{i\rightarrow j}(\tau_{i}\mid\tau_{j})=
\hspace{-0.25cm}
\sum_{\{\tau_{k}\}_{k\in\partial i\backslash j}}
\hspace{-0.34cm}
W_{RFIM}
\hspace{-0.15cm}
\prod_{k\in\partial i\backslash j}m^{k\rightarrow i}(\tau_{k}\mid\tau_{i}),
\label{eq:message_RFIM_one_app}
\end{equation}
where
\begin{equation}
W_{RFIM}=\prod_{t'=0}^{\tau_{i}-2}\left(1-\frac{1}{\tau}\mathds{1}[\Delta_i^{t'}>0]\right)\frac{1}{\tau}\mathds{1}[\Delta_i^{\tau_{i}-1}>0].
\end{equation}
Recall that here and in what follows we use a convention
\begin{equation}
\prod_{t=a}^{a-\epsilon}(\ldots) \equiv 1
\end{equation}
for any fixed $a$ and $\epsilon > 0$. We define the marginal probabilities in the cavity dynamics $D_{j}$ as
\begin{align}
p^{i\rightarrow j}(t)=
\sum_{\tau_{i} > t}m^{i\rightarrow j}(\tau_{i}\mid T), \label{eq:p_definition_app}
\\
q^{i\rightarrow j}(t)=
\sum_{\tau_{i} \leq t}m^{i\rightarrow j}(\tau_{i}\mid T),
\label{eq:q_definition_app}
\end{align}
and the probabilities of interest in the original dynamics as
\begin{align}
p^{i}(t)=
\sum_{\tau_{i} > t}m^{i}(\tau_{i}), \label{eq:p_marginal_definition_app}
\\
q^{i}(t)=1-p^{i}(t).
\label{eq:q_marginal_definition_app} 
\end{align}
The corresponding DMP equation for this model can be derived, starting from the equation (\ref{eq:message_RFIM_one_app}) and using elementary properties of the messages, such as normalization and causality constraints. First of all, let us rewrite (\ref{eq:message_RFIM_one_app}) in an equivalent form, explicitly using the monotonicity of the local field \eqref{eq:RFIM_local_field}

\begin{equation}
m^{i\rightarrow j}(\tau_{i}\mid\tau_{j})=
\frac{1}{\tau}
\sum_{\Theta_{i}^{*}=0}^{\tau_{i}-1}
\left(
1-\frac{1}{\tau}
\right)^{(\tau_{i}-1-\tau_{i}^{*})}
\sum_{\{\tau_{k}\}_{k\in\partial i\backslash j}}
\mathds{1}[\Delta_i^{\tau_{i}^{*}-1}<0]
\mathds{1}[\Delta_i^{\tau_{i}^{*}}>0]
\prod_{k\in\partial i\backslash j}m^{k\rightarrow i}(\tau_{k}\mid\tau_{i}),
\end{equation}
where $\tau_{i}^{*}$ appears explicitly as the moment when the local field becomes positive for the first time and the convention $\mathds{1}[\Delta_i^{-1}<0]\equiv 1$ is used.
In what follows, we use two natural properties of messages.

\textbf{Property 1.} The norm conservation. For every fixed $\tau_{j}$
\begin{equation}
\sum_{\tau_{i}=0}^{T}m^{i\rightarrow j}(\tau_{i}\mid\tau_{j})=1.
\label{eq:RFIM_message_property1}
\end{equation}

\textbf{Property 2.} If $\tau_{j}>\tau_{i},$ then for every $t'>\tau_{i}$
\begin{equation}
m^{i\rightarrow j}(\tau_{i}\mid \tau_{j})=m^{i\rightarrow j}(\tau_{i}\mid t').
\label{eq:RFIM_message_property2}
\end{equation}

Using the definition (\ref{eq:p_definition_app}), we get
\begin{equation}
p^{i\rightarrow j}(t+1)=
\sum_{\tau_{i} \geq t+1}
\sum_{\tau_{i}^{*}=0}^{\tau_{i}-1}
\frac{1}{\tau}
\left(
1-\frac{1}{\tau}
\right)^{(\tau_{i}-1-\tau_{i}^{*})}
\sum_{\{\Theta_{k}\}_{k\in\partial i\backslash j}}
\mathds{1}[\Delta_i^{\tau_{i}^{*}-1}<0]
\mathds{1}[\Delta_i^{\tau_{i}^{*}}>0]
\prod_{k\in\partial i\backslash j}m^{k\rightarrow i}(\tau_{k}\mid \tau_{i}).
\label{eq:RFIM_p_expression1}
\end{equation}
An important observation that can be made on this expression is that nothing changes if we replace in the right-hand side of the equation above the messages $m^{k\rightarrow i}(\tau_{k}\mid \tau_{i})$, conditioned on the flipping time $\tau_{i}$, by the messages conditioned on the stopping time $T$, $m^{k\rightarrow i}(\tau_{k}\mid T)$. The easiest way to see it consists in observing that the value of the probability $p^{i\rightarrow j}(t+1)$ should not depend on the value of the stopping time provided that $T>t+1$, and can be assigned to an arbitrary value. Since in \eqref{eq:RFIM_p_expression1} we are only interested in $\tau_{i}>t+1$, we can in particular choose $T=t+2$, and since the stopping time by definition comprises all the events that happen after the time $T$, we get
\begin{equation}
p^{i\rightarrow j}(t+1)=
\sum_{\tau_{i} \geq t+1}
\sum_{\tau_{i}^{*}=0}^{\tau_{i}-1}
\frac{1}{\tau}
\left(
1-\frac{1}{\tau}
\right)^{(\tau_{i}-1-\tau_{i}^{*})}
\sum_{\{\tau_{k}\}_{k\in\partial i\backslash j}}
\mathds{1}[\Delta_i^{\tau_{i}^{*}-1}<0]
\mathds{1}[\Delta_i^{\tau_{i}^{*}}>0]
\prod_{k\in\partial i\backslash j}m^{k\rightarrow i}(\tau_{k}\mid T).
\label{eq:RFIM_p_expression2}
\end{equation}  
The sums over $\tau_{i}$ and $\tau_{i}^{*}$ in this expression can be split into two terms:
\begin{equation}
\sum_{\tau_{i}>t+1}
\sum_{\tau_{i}^{*}=0}^{\tau_{i}-1}=
\sum_{\tau_{i}>t+1}
\delta_{\tau_{i}^{*},\tau_{i}-1}+
\sum_{\tau_{i}>t+1}
\sum_{\tau_{i}^{*}=0}^{\tau_{i}-2},
\label{eq:sum_separation}
\end{equation}
and we get
\begin{align}
\notag
p^{i\rightarrow j}(t+1)&=\sum_{\tau_{i}>t+1}\frac{1}{\tau}
\sum_{\{\tau_{k}\}_{k\in\partial i\backslash j}}
\mathds{1}[\Delta_i^{\tau_{i}-2}<0]
\mathds{1}[\Delta_i^{\tau_{i}-1}>0]
\prod_{k\in\partial i\backslash j}m^{k\rightarrow i}(\tau_{k}\mid T)
\\
\notag
&+\sum_{\tau_{i}>t+1}
\sum_{\tau_{i}^{*}=0}^{\tau_{i}-2}
\frac{1}{\tau}
\left(
1-\frac{1}{\tau}
\right)^{(\tau_{i}-1-\tau_{i}^{*})}
\sum_{\{\tau_{k}\}_{k\in\partial i\backslash j}}
\mathds{1}[\Delta_i^{\tau_{i}^{*}-1}<0]
\mathds{1}[\Delta_i^{\tau_{i}^{*}}>0]
\prod_{k\in\partial i\backslash j}m^{k\rightarrow i}(\tau_{k}\mid T)
\\
\notag
&=\frac{1}{\tau}
\sum_{\{\tau_{k}\}_{k\in\partial i\backslash j}}
\mathds{1}[\Delta_i^{t}<0]
\prod_{k\in\partial i\backslash j}m^{k\rightarrow i}(\tau_{k}\mid T)
\\
\notag
&+(1-\frac{1}{\tau})
\sum_{\tau'_{i}>t+1}
\sum_{\tau_{i}^{*}=0}^{\tau'_{i}-1}
\frac{1}{\tau}
\left(
1-\frac{1}{\tau}
\right)^{(\tau'_{i}-1-\tau_{i}^{*})}
\sum_{\{\tau_{k}\}_{k\in\partial i\backslash j}}
\mathds{1}[\Delta_i^{\tau_{i}^{*}-1}<0]
\mathds{1}[\Delta_i^{\tau_{i}^{*}}>0]
\prod_{k\in\partial i\backslash j}m^{k\rightarrow i}(\tau_{k}\mid T)
\\
&=\frac{1}{\tau}\left(1-\sum_{\{\tau_{k}\}_{k\in\partial i\backslash j}}
\mathds{1}[\Delta_i^{t}>0]
\prod_{k\in\partial i\backslash j}m^{k\rightarrow i}(\tau_{k}\mid T)\right)
+\left(1-\frac{1}{\tau}\right)p^{i\rightarrow j}(t).
\label{eq:RFIM_p_expression3}
\end{align}
The remaining sum over flipping times $\{\tau_{k}\}_{k\in\partial i\backslash j}$ can be further simplified by pushing the sum over the indicator function:
\begin{align}
\notag
&\sum_{\{\tau_{k}\}_{k\in\partial i\backslash j}}
\mathds{1}[\Delta_i^{t}>0]
\prod_{k\in\partial i\backslash j}m^{k\rightarrow i}(\tau_{k}\mid T)
\\
\notag
&=
\sum_{\{\tau_{k}\}_{k\in\partial i\backslash j}}
\left[
\sum_{\{\sigma_{k}^{t}\}_{k\in\partial i\backslash j}=\pm 1}
\prod_{k\in\partial i\backslash j}\delta\left(\frac{\sigma_{k}^{t}+1}{2},\mathds{1}[\tau_{k}<t]\right)
\right]
\mathds{1}[\Delta_i^{t}>0]
\prod_{k\in\partial i\backslash j}m^{k\rightarrow i}(\tau_{k}\mid T)
\\
\notag
&=
\sum_{ \{\sigma_{k}^{t}\}_{k\in\partial i\backslash j}=\pm 1}
\mathds{1}\left[h+h_{i}+\sum_{k\in\partial i\backslash j}J_{ki}\sigma_{k}^{t}-J_{ji}>0\right]
\prod_{k\in\partial i\backslash j}
\sum_{\{\tau_{k}\}_{k\in\partial i\backslash j}}
m^{k\rightarrow i}(\tau_{k}\mid T)\delta\left(\frac{\sigma_{k}^{t}+1}{2},\mathds{1}[\tau_{k}<t]\right)
\\
&=
\sum_{ \{\sigma_{k}^{t}\}_{k\in\partial i\backslash j}=\pm 1}
\mathds{1}\left[h+h_{i}+\sum_{k\in\partial i\backslash j}J_{ki}\sigma_{k}^{t}-J_{ji}>0\right]
\prod_{k\in\partial i\backslash j}
\prod_{k\in\partial i\backslash j:\sigma_{k}^{t}=+1}
\left[1-p^{k\rightarrow i}(t)\right]
\prod_{k\in\partial i\backslash j:\sigma_{k}^{t}=-1}
p^{k\rightarrow i}(t).
\label{eq:RFIM_pushing_sum}
\end{align}
Finally, using \eqref{eq:RFIM_pushing_sum} and rewriting \eqref{eq:RFIM_p_expression3} in terms of $q^{i\rightarrow j}(t)=1-p^{i\rightarrow j}(t)$, we recover the DMP equations \eqref{eq:Exact_dynamic_RFIM} and \eqref{eq:RFIM_marginal}.

\section{Derivation of the DMP equations for the generalized SI model}
\label{app:GSIdetails}

The dynamic belief propagation equation (\ref{eq:DBP_conditional}) for $\tau_{i} < T$ takes the following form in the generalized SI model:

\begin{equation}
m^{i\rightarrow j}(\tau_{i}\mid\tau_{j})=
\hspace{-0.12cm}
\sum_{\{\tau_{k}\}_{k\in\partial i\backslash j}}
\hspace{-0.24cm}
W_{SI}
\hspace{-0.15cm}
\prod_{k\in\partial i\backslash j}
m^{k\rightarrow i}(\tau_{k}\mid\tau_{i}),
\label{eq:message_passing_eq_SI_app}
\end{equation}
where the dynamic kernel $W_{SI}$ for the generalized SI model has the following form: 
\begin{align}
\notag W_{SI}=
P_{I}^{i}(0)\mathds{1}[\tau_{i}=0]+P_{S}^{i}(0)\mathds{1}[\tau_{i}>0]
&\prod_{t'=0}^{\tau_{i}-2}(1-\nu_{i})
\prod_{k\in\partial i}
\left(1-\epsilon_{ki}\mathds{1}[\tau_{k} \geq t'+1]\right)
\left(1-\lambda_{ki}\mathds{1}[t' \geq \tau_{k}]\right)
\\
& \times
\left(
1-(1-\nu_{i})\prod_{k\in\partial i}
\left(1-\epsilon_{ki}\mathds{1}[\tau_{k} \geq \tau_{i}]\right)
\left(1-\lambda_{ki}\mathds{1}[\tau_{i} \geq \tau_{k}+1]\right)
\right).
\end{align}

As it is described in the main text, the messages $m^{i\rightarrow j}(\tau_{i}\mid\tau_{j})$ allow one to define the marginal probabilities describing the dynamics of the SI model:
\begin{align}
& P_{S}^{i}(t)=\sum_{\tau_{i}>t} m^{i}(\tau_{i}),
\\
& P_{I}^{i}(t)=1-P_{S}^{i}(t).
\end{align}
It would also be useful to define the marginal probability that the node $i$ is in the state $S$ at a given time in the cavity dynamics $D_{j}$, in which the node $j$ is fixed to the state $S$ for all times:
\begin{equation}
P_{S}^{i \rightarrow j}(t)=\sum_{\tau_{i}>t} m^{i\rightarrow j}(\tau_{i}\mid T).
\label{eq:SI_S_def_app}
\end{equation}
By analogy with \eqref{eq:RFIM_message_property1} and \eqref{eq:RFIM_message_property2}, the messages in the dynamic cavity equation (\ref{eq:message_passing_eq_SI_app}) have the normalization and causality properties.

\textbf{Property 1.} For every fixed $\tau_{j}$ 
\begin{equation}
\sum_{\tau_{i}=0}^{T}m^{i\rightarrow j}(\tau_{i}\mid\tau_{j})=1.
\label{eq:SI_message_property1}
\end{equation}

\textbf{Property 2.} If $\tau_{j}>\tau_{i},$ then for every $t'>\tau_{i}$
\begin{equation}
m^{i\rightarrow j}(\tau_{i}\mid\tau_{j})=m^{i\rightarrow j}(\tau_{i}\mid t').
\label{eq:SI_message_property2}
\end{equation}

Using the definition (\ref{eq:SI_S_def_app}), we get for $t>0$
\begin{align}
\notag P_{S}^{i \rightarrow j}(t+1)&=P_{S}^{i}(0)\sum_{\tau_{i}>t+1} 
\sum_{\{\tau_{k}\}_{k\in\partial i\backslash j}}\mathds{1}[\tau_{j}=T]
\prod_{t'=0}^{\tau_{i}-2}
(1-\nu_{i})
\prod_{k\in\partial i}
\left(1-\epsilon_{ki}\mathds{1}[\tau_{k} \geq t'+1]\right)
\left(1-\lambda_{ki}\mathds{1}[t' \geq \tau_{k}]\right)
\\
&\times
\left(
1-(1-\nu_{i})\prod_{k\in\partial i}
\left(1-\epsilon_{ki}\mathds{1}[\tau_{k} \geq \tau_{i}]\right)
\left(1-\lambda_{ki}\mathds{1}[\tau_{i} \geq \tau_{k}+1]\right)
\right)
\prod_{k\in\partial i\backslash j}
m^{k \rightarrow i}(\tau_{k}\mid \tau_{i}).
\end{align}

Using the same arguments as in the derivation of the DMP equations for the zero-temperature RFIM, it can be shown that one can replace $\prod_{k\in\partial i\backslash j}m^{k\rightarrow i}(\tau_{k} \mid \tau_{i})$ in the right-hand side of the last expression by $\prod_{k\in\partial i\backslash j}m^{k\rightarrow i}(\tau_{k} \mid T)$ for arbitrary value of the stopping time $T>t+1$. Since we are interested in the messages $m^{i\rightarrow j}(\tau_{i} \mid T)$ with $\tau_{i}>t+1$, we have
\begin{align}
\notag P_{S}^{i \rightarrow j}(t+1)=
P_{S}^{i}(0)(1-\nu_{i})^{t+1} & (1-\epsilon_{ji})^{t+1}
\\
\notag \times
\sum_{\{\tau_{k}\}_{k\in\partial i\backslash j}}
\prod_{k\in\partial i \backslash j}
& \prod_{t'=0}^{t}
\left(1-\epsilon_{ki}\mathds{1}[\tau_{k} \geq t'+1]\right)
\left(1-\lambda_{ki}\mathds{1}[t' \geq \tau_{k}]\right)
m^{k\rightarrow i}(\tau_{k} \mid T)
\\
\notag \times
\Bigg[
\sum_{\tau_{i}>t+1}&\mathds{1}[\tau_{j}=T]
 \prod_{t'=t+1}^{\tau_{i}-2}
(1-\nu_{i})
\prod_{k\in\partial i}
\left(1-\epsilon_{ki}\mathds{1}[\tau_{k} \geq t'+1]\right)
\left(1-\lambda_{ki}\mathds{1}[t' \geq \tau_{k}]\right)
\\
& \times 
\left(
1-(1-\nu_{i})\prod_{k\in\partial i}
\left(1-\epsilon_{ki}\mathds{1}[\tau_{k} \geq \tau_{i}]\right)
\left(1-\lambda_{ki}\mathds{1}[\tau_{i} \geq \tau_{k}+1]\right)
\right)
\Bigg].
\end{align}
Developing the sum in the square brackets, one may ascertain that it gives exactly one:
\begin{align}
\notag
1-(1-\nu_{i})(1-\epsilon_{ji})&\prod_{k\in\partial i\backslash j}\left(1-\epsilon_{ki}\mathds{1}[\tau_{k} \geq t+2]\right)\left(1-\lambda_{ki}\mathds{1}[t+2 \geq \tau_{k}+1]\right)\\
\notag
+(1-\nu_{i})(1-\epsilon_{ji})&\prod_{k\in\partial i\backslash j}\left(1-\epsilon_{ki}\mathds{1}[\tau_{k} \geq t+2]\right)\left(1-\lambda_{ki}\mathds{1}[t+1 \geq \tau_{k}]\right)\\
\notag
-(1-\nu_{i})^{2}(1-\epsilon_{ji})^{2}&\prod_{k\in\partial i\backslash j}\Big[\left(1-\epsilon_{ki}\mathds{1}[\tau_{k} \geq t+2]\right)\left(1-\lambda_{ki}\mathds{1}[t+2 \geq \tau_{k}+1]\right)\Big]\\
\notag
\times &\prod_{k\in\partial i\backslash j}\Big[\left(1-\epsilon_{ki}\mathds{1}[\tau_{k} \geq t+3]\right)\left(1-\lambda_{ki}\mathds{1}[t+3 \geq \tau_{k}+1]\right)\Big]\\
\notag
+(1-\nu_{i})^{2}(1-\epsilon_{ji})^{2}&\prod_{k\in\partial i\backslash j}\Big[\left(1-\epsilon_{ki}\mathds{1}[\tau_{k} \geq t+2]\right)\left(1-\lambda_{ki}\mathds{1}[t+1 \geq \tau_{k}]\right)\Big]\\
\times &\prod_{k\in\partial i\backslash j}\Big[\left(1-\epsilon_{ki}\mathds{1}[\tau_{k} \geq t+3]\right)\left(1-\lambda_{ki}\mathds{1}[t+2 \geq \tau_{k}]\right)\Big]+\ldots = 1,
\end{align}
and therefore we obtain the factorized expression (\ref{eq:SIequations:S})
\begin{equation}
P_{S}^{i \rightarrow j}(t+1)=P_{S}^{i}(0)(1-\epsilon_{ji})^{t+1}(1 - \nu_{i})^{t+1}\prod_{k\in \partial i \backslash k}\theta^{k \rightarrow i}(t+1),
\end{equation}
where $\theta^{k \rightarrow i}(t+1)$ are given by
\begin{align}
\theta^{k \rightarrow i}(t+1)=
\sum_{\tau_{k}}
\prod_{t'=0}^{t}
\left(1-\epsilon_{ki}\mathds{1}[\tau_{k} \geq t'+1]\right)
\left(1-\lambda_{ki}\mathds{1}[t' \geq \tau_{k}]\right)
m^{k\rightarrow i}(\tau_{k} \mid T).
\end{align}
In order to close the equations on $P_{S}^{k \rightarrow i}(t),$ we recover the computational scheme for $\theta^{k \rightarrow i}(t+1):$
\begin{align}
& \theta^{k \rightarrow i}(t+1)-\theta^{k \rightarrow i}(t)
\\
& = \sum_{\tau_{k}}
\left(
\prod_{t'=0}^{t-1}
\left(1-\epsilon_{ki}\mathds{1}[\tau_{k} \geq t'+1]\right)
\left(1-\lambda_{ki}\mathds{1}[t' \geq \tau_{k}]\right)
\right)
\left(
-\epsilon_{ki}\mathds{1}[\tau_{k} \geq t+1]-\lambda_{ki}\mathds{1}[t \geq \tau_{k}]
\right)
m^{k\rightarrow i}(\tau_{k} \mid T)
\\
&=-\epsilon_{ki}\sum_{\tau_{k}}\left(
\prod_{t'=0}^{t-1}
\left(1-\epsilon_{ki}\mathds{1}[\tau_{k} \geq t'+1]\right)
\right)
m^{k\rightarrow i}(\tau_{k} \mid T)
\mathds{1}[\tau_{k} \geq t+1]
\\
&-\lambda_{ki}\sum_{\tau_{k}}
\left(
\prod_{t'=0}^{t-1}
\left(1-\epsilon_{ki}\mathds{1}[\tau_{k} \geq t'+1]\right)
\left(1-\lambda_{ki}\mathds{1}[t' \geq \tau_{k}]\right)
\right)
m^{k\rightarrow i}(\tau_{k} \mid T)
\mathds{1}[t \geq \tau_{k}]
\\
& \equiv
-\epsilon_{ki}\phi_{1}^{k \rightarrow i}(t)-\lambda_{ki}\phi_{2}^{k \rightarrow i}(t),
\end{align}
where, using the identity $\mathds{1}[\tau_{k} \geq t+1] = \mathds{1}[\tau_{k} \geq t] - \delta(\tau_{k},t)$, we get for $\phi_{1}^{k \rightarrow i}(t)$
\begin{align}
& \phi_{1}^{k \rightarrow i}(t)=
\sum_{\tau_{k}}\left(
\prod_{t'=0}^{t-2}
\left(1-\epsilon_{ki}\mathds{1}[\tau_{k} \geq t'+1]\right)
\right)
\left(1-\epsilon_{ki}\mathds{1}[\tau_{k} \geq t]\right)
m^{k\rightarrow i}(\tau_{k} \mid T)
\mathds{1}[\tau_{k} \geq t+1]
\\
& = (1-\epsilon_{ki})\phi_{1}^{k \rightarrow i}(t-1)-
\left(\prod_{t'=0}^{t-1}(1-\epsilon_{ki})\right)m^{k\rightarrow i}(t \mid T)=
(1-\epsilon_{ki})\phi_{1}^{k \rightarrow i}(t-1)-
(1-\epsilon_{ki})^{t}(P_{S}^{k \rightarrow i}(t-1)-P_{S}^{k \rightarrow i}(t)),
\end{align}
and for $\phi_{2}^{k \rightarrow i}(t)$
\begin{align}
& \phi_{2}^{k \rightarrow i}(t)=
\sum_{\tau_{k}}\left(
\prod_{t'=0}^{t-2}
\left(1-\epsilon_{ki}\mathds{1}[\tau_{k} \geq t'+1]\right)
\left(1-\lambda_{ki}\mathds{1}[t' \geq \tau_{k}]\right)
\right)
\left(1-\epsilon_{ki}\mathds{1}[\tau_{k} \geq t]\right)
\\
& \times
\left(1-\lambda_{ki}\mathds{1}[t \geq \tau_{k}+1]\right)
m^{k\rightarrow i}(\tau_{k} \mid T)
\mathds{1}[t \geq \tau_{k}]=
(1-\lambda_{ki})\phi_{2}^{k \rightarrow i}(t-1)+
(1-\epsilon_{ki})^{t}(P_{S}^{k \rightarrow i}(t-1)-P_{S}^{k \rightarrow i}(t)).
\end{align}
This gives exactly the computational scheme (\ref{eq:SIequations:S}-\ref{eq:SIequations:phi2}).

\section{Derivation of the DMP equations for the SIR model}
\label{app:SIRdetails}

The derivation of the DMP equations for the SIR model follows the very same lines as the derivation for the SI model, but includes several subtleties. First of all, since now each node can be in one of three states ($S$ - susceptible, $I$ - infected and $R$ - recovered), the trajectory of the node $i$ can be parametrized by two times $\tau_{i}$ and $\omega_{i}$: $\tau_{i}$ is defined as a first time to be in the state $I,$ while $\omega_{i}$ is a first time to be in the state $R$. We will denote the trajectory of the spin $i$ as $\vec{\sigma}_{i}(t)=(\tau_{i},\omega_{i})$. Therefore, the dynamic BP equation (\ref{eq:DBP_conditional}) for $\omega_{i} < T$ becomes in this case:

\begin{equation}
m^{i\rightarrow j}(\tau_{i},\omega_{i}\mid\tau_{j},\omega_{j})=
\sum_{\{\tau_{k},\omega_{k}\}_{k\in\partial i\backslash j}} W_{SIR}
\prod_{k\in\partial i\backslash j}
m^{k\rightarrow i}(\tau_{k},\omega_{k}\mid\tau_{i},\omega_{i}),\label{eq:message_passing_eq_SIR}
\end{equation}
where

\begin{align}
\notag W_{SIR}= \Bigg[
P_S^{i}(0)\mathds{1}[\tau_{i}>0)]
& \prod_{t'=0}^{\tau_{i}-2}
\prod_{k\in\partial i}
\left(
1-\lambda_{ki}\mathds{1}[\omega_{k} \geq t'+1]\mathds{1}[\tau_{k} \leq t']
\right)
\left(
1-\prod_{k\in\partial i}\left( 1-\lambda_{ki}\mathds{1}[\omega_{k} \geq \tau_{i}]\mathds{1}[\tau_{k} \leq \tau_{i}-1]\right)
\right)
\\
& + P_I^{i}(0)\mathds{1}[\tau_{i}=0]
\Bigg] \times
\left(
\prod_{t''=\tau_{i}}^{\omega_{i}-2}
(1-\mu_{i})
\right)
\times \mu_{i}
\times \prod_{k\in\partial i}
\mathds{1}[\omega_{k} \geq \tau_{k}+1]
\mathds{1}[\omega_{i} \geq \tau_{i}+1].
\end{align}
The marginals of interest in the SIR model can be defined as
\begin{align}
& P_{S}^{i}(t)=\sum_{\tau_{i}>t} \: \sum_{\omega_{i}>\tau_{i}}m^{i}(\tau_{i},\omega_{i}),
\label{eq:SIR_S_def_marginal}
\\
& P_{I}^{i}(t)=\sum_{\tau_{i} \leq t} \: \sum_{\omega_{i}>t}m^{i}(\tau_{i},\omega_{i}),
\label{eq:SIR_I_def_marginal}
\\
& P_{R}^{i}(t)=\sum_{\omega_{i} \leq t} \: \sum_{ \tau_{i} < \omega_{i} }m^{i}(\tau_{i}, \omega_{i}).
\label{eq:SIR_R_def_marginal}
\end{align}
We also define the marginal probability for the susceptible state in the corresponding cavity graph:
\begin{equation}
P_{S}^{i \rightarrow j}(t)=\sum_{\tau_{i}>t} \: \sum_{\omega_{i}>\tau_{i}}m^{i\rightarrow j}(\tau_{i},\omega_{i}\mid T,T).
\label{eq:SIR_S_def}
\end{equation}
Let us point out the properties of the messages.

\textbf{Property 1.} $m^{i\rightarrow j}(\tau_{i}, \omega_{i} \mid T,T)=0$ if $\tau_{i} \geq \omega_{i}$;

\textbf{Property 2.} If $\tau_{j} \geq \tau_{i},$ then $m^{i\rightarrow j}(\tau_{i}, \omega_{i} \mid \tau_{j}, \omega_{j})=
m^{i\rightarrow j}(\tau_{i}, \omega_{i} \mid t', \omega_{j})$ for every $\tau_{i} \leq t' < \omega_{j}$;

\textbf{Property 3.} $\sum_{\tau_{i}, \omega_{i}}m^{i\rightarrow j}(\tau_{i}, \omega_{i} \mid T,T)=1$;

\textbf{Property 4.} $m^{i\rightarrow j}(\tau_{i}, \omega_{i}+1 \mid T,T)=(1-\mu_{i})m^{i\rightarrow j}(\tau_{i}, \omega_{i} \mid T,T)$.

The properties, equivalent to properties 1, 2 and 4, are also valid for marginals $m^{i}(\tau_{i}, \omega_{i})$. It is straightforward to establish first two evolution equations on the quantities $P_{S}^{i}(t)$, $P_{I}^{i}{t}$ and  $P_{R}^{i}(t)$. According to the definitions,
\begin{equation}
P_{R}^{i}(t+1) = \sum_{\omega_{i} \leq t+1} \: \sum_{ \tau_{i} < \omega_{i} }
m^{i}(\tau_{i}, \omega_{i}) =
\sum_{\omega_{i} \leq t} \: \sum_{ \tau_{i} < \omega_{i} }
m^{i}(\tau_{i}, \omega_{i})
+\delta_{\omega_{i},t+1}\sum_{\tau_{i} \leq t}
m^{i}(\tau_{i} \omega_{i})
=P_{R}^{i}(t)+\mu_{i}P_{I}^{i}(t), \label{eq:SIR_R_equation}
\end{equation}
where we used the property 4 of marginals, because
\begin{align}
\sum_{\omega_{i} \geq t+1}
m^{i}(\tau_{i}, \omega_{i})=
\frac{1}{1-(1-\mu_{i})}
m^{i}(\tau_{i}, t+1)=
\frac{1}{\mu_{i}}m^{i}(\tau_{i}, t+1).
\end{align}
Since the expressions defined in (\ref{eq:SIR_S_def_marginal}-\ref{eq:SIR_R_def_marginal}) sum to one, it is obvious that
\begin{align}
P_{I}^{i}(t+1)=1-P_{S}^{i}(t+1)-P_{R}^{i}(t+1). \label{eq:SIR_I_equation}
\end{align}
In what follows we show that we can put $P_{S}^{i \rightarrow j}(t+1)$ in the form
\begin{align}
P_{S}^{i \rightarrow j}(t+1)=P_{S}^{i}(0)\prod_{k\in\partial i \backslash j}\theta^{k \rightarrow i}(t+1), \label{eq:SIR_S_equation}
\end{align}
where $\theta^{k \rightarrow i}(t+1)$ can be calculated via $P_{S}^{k \rightarrow i}(t)$ at each time step. The equations (\ref{eq:SIR_R_equation}), (\ref{eq:SIR_I_equation}), (\ref{eq:SIR_S_equation}) and its marginalized version, together with a computational scheme for $\theta^{i \rightarrow j}(t+1),$ form a closed set of equations for the SIR model.

We proceed in the same way as for the SI model. The quantity $\theta^{k \rightarrow i}(t+1)$ is now defined as
\begin{align}
\theta^{k \rightarrow i}(t+1)=
\sum_{\tau_{k},\omega_{k}}
\mathds{1}[\omega_{k} \geq \tau_{k}+1]
\prod_{t'=0}^{t}
\left(
1-\lambda_{ki}\mathds{1}[\omega_{k} \geq t'+1]\mathds{1}[\tau_{k} \leq t']
\right)
m^{k\rightarrow i}(\tau_{k},\omega_{k} \mid T,T).
\end{align}
As for the SI model, we can write
\begin{equation}
\theta^{k \rightarrow i}(t+1)-\theta^{k \rightarrow i}(t)\equiv
-\lambda_{ki}\phi^{k \rightarrow i}(t), \label{eq:SIR_theta_computational_scheme}
\end{equation}
where we get for $\phi^{k \rightarrow i}(t)$
\begin{align}
\notag & \phi^{k \rightarrow i}(t)=
\\
\notag &
=\sum_{\tau_{k},\omega_{k}>\tau_{k}}
\left(
\prod_{t'=0}^{t-2}
\left(
1-\lambda_{ki}\mathds{1}[\tau_{k} \leq t']
\right)
\right)
\left(
1-\lambda_{ki}\mathds{1}[\tau_{k} \leq t-1]
\right)
\mathds{1}[\omega_{k} \geq t+1]\mathds{1}[\tau_{k} \leq t]
m^{k\rightarrow i}(\tau_{k},\omega_{k} \mid T,T)
\\
\notag &
=(1-\lambda_{ki})(1-\mu_{k})\phi^{k \rightarrow i}(t-1)-
\sum_{\omega_{k}}
\mathds{1}[\omega_{k} \geq t+1]
m^{k\rightarrow i}(t,\omega_{k} \mid T,T)
\\
&
=(1-\lambda_{ki})(1-\mu_{k})\phi^{k \rightarrow i}(t-1)-
(P_{S}^{k \rightarrow i}(t)-P_{S}^{k \rightarrow i}(t-1)). \label{eq:SIR_phi_computational_scheme}
\end{align}
In the last expression the factor $(1-\mu_{k})$ is due to the property 4 of messages. Equations (\ref{eq:SIR_theta_computational_scheme}) and (\ref{eq:SIR_phi_computational_scheme}) complete the computational scheme for $\theta^{k \rightarrow i}(t+1)$ for each time step, and we recover the DMP equations (\ref{eq:SIRequations:P_S}-\ref{eq:SIRequations:I}).

\section{Derivation of the DMP equations for the rumor spreading model}
\label{app:Rumordetails}

Let us follow the main lines of derivation of these DMP equations, concentrating mainly on the subtleties with the respect to the SIR model. Again, the trajectory of the node $i$ can be parametrized by two flipping times: $\tau_{i}$ (first time in $I$) and $\omega_{i}$ (first time in $R$). The trajectory of the spin $i$ is hence described by $\vec{\sigma}_{i}(t)=(\tau_{i},\omega_{i})$. The dynamic BP equation (\ref{eq:DBP_conditional}) in the rumor spreading model reads for $\omega_{i} < T$:

\begin{equation}
m^{i\rightarrow j}(\tau_{i},\omega_{i}\mid\tau_{j},\omega_{j})=
\sum_{\{\tau_{k},\omega_{k}\}_{k\in\partial i\backslash j}} W_{RS}
\prod_{k\in\partial i\backslash j}
m^{k\rightarrow i}(\tau_{k},\omega_{k}\mid\tau_{i},\omega_{i}),\label{eq:message_passing_eq_RS}
\end{equation}
where

\begin{align}
\notag W_{RS}= \Bigg[
P_{S}^{i}(0)\mathds{1}[\tau_{i}>0)]
\prod_{t'=0}^{\tau_{i}-2}
\prod_{k\in\partial i}
\left(
1-\lambda_{ki}\mathds{1}[\omega_{k} \geq t'+1]\mathds{1}[\tau_{k} \leq t']
\right)
\left(
1-\prod_{k\in\partial i}\left( 1-\lambda_{ki}\mathds{1}[\omega_{k} \geq \tau_{i}]\mathds{1}[\tau_{k} \leq \tau_{i}-1]\right)
\right)+ & \\
\notag
+ P_I^{i}(0)\mathds{1}[\tau_{i}=0]
\Bigg] \times
\prod_{t''=\tau_{i}}^{\omega_{i}-2}
\prod_{k\in\partial i}
\left(
1-\alpha_{ki}\mathds{1}[\omega_{k} \geq t''+1]\mathds{1}[\tau_{k} \leq t'']
\right)
\left(
1-\prod_{k\in\partial i}\left(1-\alpha_{ki}\mathds{1}[\omega_{k} \geq \omega_{i}]\mathds{1}[\tau_{k} \leq \omega_{i}-1]\right)
\right) \times & \\
\times \prod_{k\in\partial i}
\mathds{1}[\omega_{k} \geq \tau_{k}+1]
\mathds{1}[\omega_{i} \geq \tau_{i}+1] & .
\label{eq:dynamics_RS}
\end{align}

Quantities $P_{S}^{i}(t)$, $P_{I}^{i}(t)$, $P_{R}^{i}(t)$ and $P_{S}^{i \rightarrow j}(t)$ are defined in the same way as in (\ref{eq:SIR_S_def_marginal}-\ref{eq:SIR_S_def}).

The following properties of the messages hold.

\textbf{Property 1.} $m^{i\rightarrow j}(\tau_{i}, \omega_{i} \mid T,T)=0$ if $\tau_{i} \geq \omega_{i}$;

\textbf{Property 2.} If $\tau_{j} \geq \tau_{i},$ then $m^{i\rightarrow j}(\tau_{i}, \omega_{i} \mid \tau_{j}, \omega_{j})=
m^{i\rightarrow j}(\tau_{i}, \omega_{i} \mid t', \omega_{j})$ for every $\tau_{i} \leq t' < \omega_{j}$;

\textbf{Property 3.} $\sum_{\tau_{i}, \omega_{i}}m^{i\rightarrow j}(\tau_{i}, \omega_{i} \mid T,T)=1$.

According to the definitions, we have
\begin{equation}
P_{R}^{i}(t+1) = \sum_{\omega_{i} \leq t+1} \: \sum_{ \tau_{i} < \omega_{i} }
m^{i\rightarrow j}(\tau_{i}, \omega_{i}) =
\sum_{\omega_{i} \leq t} \: \sum_{ \tau_{i} < \omega_{i} }
m^{i}(\tau_{i}, \omega_{i})
+\delta_{\omega_{i},t+1}\sum_{\tau_{i} \leq t}
m^{i}(\tau_{i}, \omega_{i})
=P_{R}^{i}(t)+\sum_{\tau_{i} \leq t}
m^{i}(\tau_{i}, \omega_{i}). \label{eq:RS_R_equation}
\end{equation}
The expressions defined in (\ref{eq:SIR_S_def_marginal}-\ref{eq:SIR_R_def_marginal}) sum to one, hence
\begin{align}
P_{I}^{i}(t+1)=1-P_{S}^{i}(t+1)-P_{R}^{i}(t+1). \label{eq:RS_I_equation}
\end{align}
As for the SIR model, we show that we can put $P_{S}^{i \rightarrow j}(t+1)$ in the form
\begin{align}
P_{S}^{i \rightarrow j}(t+1)=P_{S}^{i}(0)\prod_{k\in\partial i \backslash j}\theta^{k \rightarrow i}(t+1), \label{eq:RS_S_equation}
\end{align}
where $\theta^{k \rightarrow i}(t+1)$ is defined as
\begin{align}
\theta^{k \rightarrow i}(t+1)=
\sum_{\tau_{k},\omega_{k}}
\mathds{1}[\omega_{k} \geq \tau_{k}+1]
\prod_{t'=0}^{t}
\left(
1-\lambda_{ki}\mathds{1}[\omega_{k} \geq t'+1]\mathds{1}[\tau_{k} \leq t']
\right)
m^{k\rightarrow i}(\tau_{k},\omega_{k} \mid T,T),
\end{align}
and we have in the same way
\begin{equation}
\theta^{k \rightarrow i}(t+1)-\theta^{k \rightarrow i}(t)\equiv
-\lambda_{ki}\phi^{k \rightarrow i}(t). \label{eq:RS_theta_computational_scheme}
\end{equation}
The quantity $\phi^{k \rightarrow i}(t)$ is defined as
\begin{equation}
\phi^{k \rightarrow i}(t)=
\sum_{\tau_{k},\omega_{k}>\tau_{k}}
\left(
\prod_{t'=0}^{t-1}
\left(
1-\lambda_{ki}\mathds{1}[\tau_{k} \leq t']
\right)
\right)
\mathds{1}[\omega_{k} \geq t+1]\mathds{1}[\tau_{k} \leq t]
m^{k\rightarrow i}(\tau_{k},\omega_{k} \mid T,T).
\end{equation}
Since $\mathds{1}[\omega_{k} \geq t+1]=\mathds{1}[\omega_{k} \geq t]-\delta_{\omega_{k},t}$ and $\mathds{1}[\tau_{k} \leq t]=\mathds{1}[\tau_{k} \leq t-1]+\delta_{\tau_{k},t}$, we obtain
\begin{equation}
\phi^{k \rightarrow i}(t)=(1-\lambda_{ki})\phi^{k \rightarrow i}(t-1)+P_S^{k \rightarrow i}(t-1)-P_S^{k \rightarrow i}(t)-\sum_{\tau_{k}\leq t-1}\prod_{t'=0}^{t-1}\left(1-\lambda_{ki}\mathds{1}[\tau_{k}\leq t']\right)m^{k \rightarrow i}(\tau_{k},t\mid T,T).
\label{eq:RS_phi_computational_scheme}
\end{equation}
The last term corresponds to the probability of recovering at time step $t$. As for the SIR model, the initial conditions for $\theta^{k \rightarrow i}(t)$ and $\phi^{k \rightarrow i}(t)$ are given by $\theta^{k\rightarrow i}(0)=1$, and $\phi^{k\rightarrow i}(0)=\delta_{\sigma_{k}^{0},I}$. Since these equations are in not in a closed form, we proceed to the computation of $m^{i \rightarrow j}(\tau_{i},t\mid T,T)$ for $\tau_{i}<t$. The equations \eqref{eq:RSequations:m0} and \eqref{eq:RSequations:m} follow directly from \eqref{eq:message_passing_eq_RS} with dynamics \eqref{eq:dynamics_RS}, if we denote (for $t_{1}=t-2$ or $t_{1}=t-1$)
\begin{align}
\notag
& \chi_{1}^{k \rightarrow i}(\tau_{i}-2,t_{1})=
\\
& \hspace{-0.16cm}
\sum_{\tau_{k},\omega_{k}>\tau_{k}}
\hspace{-0.16cm}
\left(
\prod_{t'=0}^{\tau_{i}-2}
\left(
1-\lambda_{ki}\mathds{1}[\omega_{k} \geq t'+1]\mathds{1}[\tau_{k} \leq t']
\right)
\right)
\left(
\prod_{t''=\tau_{i}}^{t_{1}}
\left(
1-\alpha_{ki}\mathds{1}[\omega_{k} \geq t''+1]\mathds{1}[\tau_{k} \leq t'']
\right)
\right)
m^{k \rightarrow i}(\tau_{k},\omega_{k}\mid \tau_{i},t),
\end{align}
and
\begin{align}
\notag
& \chi_{2}^{k \rightarrow i}(\tau_{i}-1,t_{1})=
\\
& \hspace{-0.16cm}
\sum_{\tau_{k},\omega_{k}>\tau_{k}}
\hspace{-0.16cm}
\left(
\prod_{t'=0}^{\tau_{i}-1}
\left(
1-\lambda_{ki}\mathds{1}[\omega_{k} \geq t'+1]\mathds{1}[\tau_{k} \leq t']
\right)
\right)
\left(
\prod_{t''=\tau_{i}}^{t_{1}}
\left(
1-\alpha_{ki}\mathds{1}[\omega_{k} \geq t''+1]\mathds{1}[\tau_{k} \leq t'']
\right)
\right)
m^{k \rightarrow i}(\tau_{k},\omega_{k}\mid \tau_{i},t).
\end{align}
Similarly to the evolution of $\theta^{k \rightarrow i}(t)$, the evolution equations for $\chi_{1}^{k \rightarrow i}(\tau_{i}-2,t_{1})$ and $\chi_{2}^{k \rightarrow i}(\tau_{i}-1,t_{1})$, given their definitions, read
\begin{align}
\chi_{1}^{k \rightarrow i}(\tau_{i}-2,t-1) =& \chi_{1}^{k \rightarrow i}(\tau_{i}-2,t-2) 
-\alpha_{ki}\psi_{1}^{k \rightarrow i}(\tau_{i}-2,t-1),
\\
\chi_{2}^{k \rightarrow i}(\tau_{i}-1,t-1) =& \chi_{2}^{k \rightarrow i}(\tau_{i}-1,t-2) 
-\alpha_{ki}\psi_{2}^{k \rightarrow i}(\tau_{i}-1,t-1),
\end{align}
where $\psi_{1}^{k \rightarrow i}(\tau_{i}-2,t_{1})$ and $\psi_{2}^{k \rightarrow i}(\tau_{i}-1,t_{1})$ are defined as

\begin{align}
\notag
\psi_{1}^{k \rightarrow i}(\tau_{i}-2,t_{1})=
\hspace{-0.16cm}
\sum_{\tau_{k},\omega_{k}>\tau_{k}}
\hspace{-0.16cm}
&\left(
\prod_{t'=0}^{\tau_{i}-2}
\left(
1-\lambda_{ki}\mathds{1}[\omega_{k} \geq t'+1]\mathds{1}[ \tau_{k} \leq t']
\right)
\right)
\left(
\prod_{t''=\tau_{i}}^{t_{1}-1}
\left(
1-\alpha_{ki}\mathds{1}[\omega_{k} \geq t''+1]\mathds{1}[\tau_{k} \leq t'']
\right)
\right)
\\
&\times
\mathds{1}[\omega_{k} \geq t_{1}+1]\mathds{1}[\tau_{k} \leq t_{1}]
m^{k \rightarrow i}(\tau_{k},\omega_{k}\mid \tau_{i},t),
\end{align}
and
\begin{align}
\notag
\psi_{2}^{k \rightarrow i}(\tau_{i}-1,t_{1})=
\hspace{-0.16cm}
\sum_{\tau_{k},\omega_{k}>\tau_{k}}
\hspace{-0.16cm}
&\left(
\prod_{t'=0}^{\tau_{i}-1}
\left(
1-\lambda_{ki}\mathds{1}[\omega_{k} \geq t'+1]\mathds{1}[ \tau_{k} \leq t']
\right)
\right)
\left(
\prod_{t''=\tau_{i}}^{t_{1}-1}
\left(
1-\alpha_{ki}\mathds{1}[\omega_{k} \geq t''+1]\mathds{1}[\tau_{k} \leq t'']
\right)
\right)
\\
&\times
\mathds{1}[\omega_{k} \geq t_{1}+1]\mathds{1}[\tau_{k} \leq t_{1}]
m^{k \rightarrow i}(\tau_{k},\omega_{k}\mid \tau_{i},t).
\end{align}
 
In order to close the computational scheme, we have to write the update equations for $\psi_{1}^{k \rightarrow i}(\tau_{i}-2,t_{1})$ and $\psi_{2}^{k \rightarrow i}(\tau_{i}-1,t_{1})$. One has to be careful, because the messages involved in these quantities are conditioned on the state of the node $i$, with flipping times $(\tau_{i},t)$.  Using again the identities $\mathds{1}[\omega_{k} \geq t_{2}+1]=\mathds{1}[\omega_{k} \geq t_{2}]-\delta_{\omega_{k},t_{2}}$ and $\mathds{1}[\tau_{k} \leq t_{2}]=\mathds{1}[\tau_{k} \leq t_{2}-1]+\delta_{\tau_{k},t_{2}}$, we recover the equations \eqref{eq:RSequations:psi1} and \eqref{eq:RSequations:psi2}, with the definition of conditional probability of staying in the state $S$ \eqref{eq:RSequations:PS_conditioned}.

Finally one can see, directly from the definitions, that the initial conditions for this computational scheme are given by $\chi_{1}^{k \rightarrow i}(-2,-1)=1$ and $\psi_{1}^{k \rightarrow i}(-2,0)=\phi^{k \rightarrow i}(0)$, and the border conditions for $\tau_{i}=t-1$ are given for each time step by expressions (\ref{eq:RSequations:border_condition_chi1}-\ref{eq:RSequations:border_condition_psi2}).

\section{Example of derivation of the DMP equations for the $K=4$ states model with unidirectional dynamics}
\label{app:K4details}

The dynamic BP equation (\ref{eq:DBP_conditional}) in the minimal $K=4$ model reads for $\varepsilon_{i} < T$:

\begin{equation}
m^{i\rightarrow j}(\tau_{i},\omega_{i},\varepsilon_{i}\mid\tau_{j},\omega_{j},\varepsilon_{j})=
\sum_{\{\tau_{k},\omega_{k},\varepsilon_{k}\}_{k\in\partial i\backslash j}} W_{4}
\prod_{k\in\partial i\backslash j}
m^{k\rightarrow i}(\tau_{k},\omega_{k},\varepsilon_{k}\mid\tau_{i},\omega_{i},\varepsilon_{i}),\label{eq:message_passing_eq_Kmodel}
\end{equation}
where

\begin{align}
\notag
W_{4}= \Bigg[
P_{S}^{i}(0)\mathds{1}[\tau_{i}>0)]
\prod_{t'=0}^{\tau_{i}-2}
\prod_{k\in\partial i}
\left(
1-\lambda_{ki}\mathds{1}[\omega_{k} \geq t'+1]\mathds{1}[\tau_{k} \leq t']
\right)
\left(
1-\prod_{k\in\partial i}\left( 1-\lambda_{ki}\mathds{1}[\omega_{k} \geq \tau_{i}]\mathds{1}[\tau_{k} \leq \tau_{i}-1]\right)
\right) &  \\
\notag
+ P_I^{i}(0)\mathds{1}[\tau_{i}=0]
\Bigg] \times
\prod_{t''=\tau_{i}}^{\omega_{i}-2}
\prod_{k\in\partial i}
\left(
1-\alpha_{ki}\mathds{1}[\omega_{k} \geq t''+1]\mathds{1}[\tau_{k} \leq t'']
\right)
\left(
1-\prod_{k\in\partial i}\left(1-\alpha_{ki}\mathds{1}[\omega_{k} \geq \omega_{i}]\mathds{1}[\tau_{k} \leq \omega_{i}-1]\right)
\right) & \\
\notag
\times
\prod_{t'''=\omega_{i}}^{\varepsilon_{i}-2}
\prod_{k\in\partial i}
\left(
1-\beta_{ki}\mathds{1}[\varepsilon_{k} \geq t'''+1]\mathds{1}[\omega_{k} \leq t''']
\right)
\left(
1-\prod_{k\in\partial i}\left(1-\beta_{ki}\mathds{1}[\varepsilon_{k} \geq \varepsilon_{i}]\mathds{1}[\omega_{k} \leq \varepsilon_{i}-1]\right)
\right) &
\\
\times
\prod_{k\in\partial i}
\mathds{1}[\varepsilon_{k} \geq \omega_{k}+1]
\mathds{1}[\omega_{k} \geq \tau_{k}+1]
\mathds{1}[\omega_{i} \geq \tau_{i}+1]. &
\label{eq:dynamics_Kmodel}
\end{align}

Quantities $P_{S}^{i}(t)$, $P_{I_{1}}^{i}(t)$, $P_{I_{2}}^{i}(t)$, $P_{R}^{i}(t)$ and $P_{S}^{i \rightarrow j}(t)$ are defined in a way similar to (\ref{eq:SIR_S_def_marginal}-\ref{eq:SIR_S_def}). For instance,
\begin{equation}
P_{S}^{i \rightarrow j}(t)=\sum_{\tau_{i}>t} \: \sum_{\omega_{i}>\tau_{i}} \: \sum_{\varepsilon_{i}>\omega_{i}}
m^{i\rightarrow j}(\tau_{i},\omega_{i},\varepsilon_{i}\mid T,T,T).
\label{eq:K_S_def}
\end{equation}
As previously, we can put $P_{S}^{i \rightarrow j}(t+1)$ in the form
\begin{align}
P_{S}^{i \rightarrow j}(t+1)=P_{S}^{i}(0)\prod_{k\in\partial i \backslash j}\theta^{k \rightarrow i}(t+1), \label{eq:K_S_equation}
\end{align}
where $\theta^{k \rightarrow i}(t+1)$ is now defined as
\begin{align}
\theta^{k \rightarrow i}(t+1)=
\sum_{\tau_{k},\omega_{k},\varepsilon_{k}}
\mathds{1}[\varepsilon_{k} \geq \omega_{k}+1]
\mathds{1}[\omega_{k} \geq \tau_{k}+1]
\prod_{t'=0}^{t}
\left(
1-\lambda_{ki}\mathds{1}[\omega_{k} \geq t'+1]\mathds{1}[\tau_{k} \leq t']
\right)
m^{k\rightarrow i}(\tau_{k},\omega_{k},\varepsilon_{k} \mid T,T,T),
\end{align}
and we have in the same way
\begin{equation}
\theta^{k \rightarrow i}(t+1)-\theta^{k \rightarrow i}(t)\equiv
-\lambda_{ki}\phi^{k \rightarrow i}(t). \label{eq:K_theta_computational_scheme}
\end{equation}
The quantity $\phi^{k \rightarrow i}(t)$ is now defined as
\begin{equation}
\phi^{k \rightarrow i}(t)=
\sum_{\substack{\tau_{k},\omega_{k}>\tau_{k} \\ \varepsilon_{k}>\omega_{k}}}
\left(
\prod_{t'=0}^{t-1}
\left(
1-\lambda_{ki}\mathds{1}[\tau_{k} \leq t']
\right)
\right)
\mathds{1}[\omega_{k} \geq t+1]\mathds{1}[\tau_{k} \leq t]
m^{k\rightarrow i}(\tau_{k},\omega_{k},\varepsilon_{k} \mid T,T,T).
\end{equation}
Since $\mathds{1}[\omega_{k} \geq t+1]=\mathds{1}[\omega_{k} \geq t]-\delta_{\omega_{k},t}$ and $\mathds{1}[\tau_{k} \leq t]=\mathds{1}[\tau_{k} \leq t-1]+\delta_{\tau_{k},t}$, we obtain
\begin{equation}
\phi^{k \rightarrow i}(t)=(1-\lambda_{ki})\phi^{k \rightarrow i}(t-1)+P_S^{k \rightarrow i}(t-1)-P_S^{k \rightarrow i}(t)-\sum_{\tau_{k}\leq t-1}\prod_{t'=0}^{t-1}\left(1-\lambda_{ki}\mathds{1}[\tau_{k}\leq t']\right) \sum_{\varepsilon_{k} \geq t+1} m^{k \rightarrow i}(\tau_{k},t,\varepsilon_{k}\mid T,T,T).
\label{eq:K_phi_computational_scheme}
\end{equation}
The last term corresponds to the probability of switching to the $I_{2}$ state at time step $t$. We see that for this term, exactly the same problem as the one indicated in the main text for the equation \eqref{eq:Kequations:I2} remains: the sum over $\varepsilon_{k}$ goes over a number of terms of order of $T$. At the same time, we see that both equations \eqref{eq:Kequations:I2} and \eqref{eq:K_phi_computational_scheme} have a finite number of terms if we define new variables
\begin{equation}
\mu^{k \rightarrow i}(\tau_{k},t \mid T,T) = \sum_{\varepsilon_{k} \geq t+1} m^{k \rightarrow i}(\tau_{k},t,\varepsilon_{k}\mid T,T,T).
\label{eq:K_mu_def}
\end{equation}
On the other hand, in order to close the computational scheme (\ref{eq:Kequations:S}-\ref{eq:Kequations:I1}), \eqref{eq:K_S_equation} \eqref{eq:K_theta_computational_scheme}, \eqref{eq:K_phi_computational_scheme}, one needs to write the evolution equations for both $m^{i\rightarrow j}(\tau_{i},\omega_{i},t \mid T,T,T)$ and $\mu^{i \rightarrow j}(\tau_{i},t \mid T,T)$ variables. The evolution of $\mu^{i \rightarrow j}(\tau_{i},t \mid T,T)$ follows the same equations as for the rumor spreading model (\ref{eq:RSequations:m0}-\ref{eq:RSequations:border_condition_psi2}), except that now we will require the computation of $\mu^{i \rightarrow j}(\tau_{i},t \mid \tau_{j},\omega_{j})$. Since all the functions $\chi_{1}$ and $\chi_{2}$ are independent on $(\tau_{j},\omega_{j})$, we will only need to compute the equivalents of \eqref{eq:RSequations:rho1} and \eqref{eq:RSequations:rho2}. These equivalents are given by the coefficients $\hat{\rho}_{1}^{j \rightarrow i}$ and $\hat{\rho}_{2}^{j \rightarrow i}$:
\begin{align}
\hat{\rho}^{j \rightarrow i}_{1}(\tau_{i}-2,\omega_{i}-2,t_{1}\mid \tau_{j},\omega_{j})
=\prod_{t'=0}^{\tau_{i}-2}\left(1-\lambda_{ji}\mathds{1}[\tau_{j}\leq t']\mathds{1}[\omega_{j}\geq t'+1]\right)
&\prod_{t''=\tau_{i}}^{\omega_{i}-2}\left(1-\alpha_{ji}\mathds{1}[\tau_{j}\leq t'']\mathds{1}[\tau_{j}\leq t']\mathds{1}[\omega_{j}\geq t''+1]\right),
\\
\hat{\rho}^{j \rightarrow i}_{2}(\tau_{i}-2,\omega_{i}-1,t_{1}\mid \tau_{j},\omega_{j})
=\prod_{t'=0}^{\tau_{i}-2}\left(1-\lambda_{ji}\mathds{1}[\tau_{j}\leq t']\mathds{1}[\omega_{j}\geq t'+1]\right)
&\prod_{t''=\tau_{i}}^{\omega_{i}-1}\left(1-\alpha_{ji}\mathds{1}[\tau_{j}\leq t'']\mathds{1}[\tau_{j}\leq t']\mathds{1}[\omega_{j}\geq t''+1]\right),
\end{align}
which are the restrictions of $\zeta^{j \rightarrow i}_{1}$ and $\zeta^{j \rightarrow i}_{2}$ (see definitions below) to the case $\beta_{ij}=0$.

For the $m^{i\rightarrow j}(\tau_{i},\omega_{i},t \mid T,T,T)$, we proceed in the similar way as for the derivation of \eqref{eq:RSequations:m0} and \eqref{eq:RSequations:m} in the rumor spreading model, but now we will have, instead of two and four terms, correspondingly four and eight. For each $t$, directly from \eqref{eq:message_passing_eq_Kmodel} and \eqref{eq:dynamics_Kmodel}, we have
\begin{align}
\notag
m^{i \rightarrow j}(0, \omega_{i}, t\mid \tau_{j},\omega_{j},T)=P^{i}_{I}(0)\Big[
&g^{i \rightarrow j}_{\zeta_{1},\xi_{1}}(-2,\omega_{i}-2,t-2\mid \tau_{j},\omega_{j})
-g^{i \rightarrow j}_{\zeta_{1},\xi_{1}}(-2,\omega_{i}-2,t-1\mid \tau_{j},\omega_{j})
\\
&-g^{i \rightarrow j}_{\zeta_{2},\xi_{2}}(-2,\omega_{i}-1,t-2\mid \tau_{j},\omega_{j})
+g^{i \rightarrow j}_{\zeta_{2},\xi_{2}}(-2,\omega_{i}-1,t-1\mid \tau_{j},\omega_{j})
\Big]
\label{eq:Kequations:m0}
\end{align}
for $1 \leq \omega_{i} \leq t-1$, $0 \leq \tau_{j} \leq t-1$ and $1 \leq \omega_{j} \leq t$, and
\begin{align}
\notag
&m^{i \rightarrow j}(\tau_{i},\omega_{i},t\mid \tau_{j},\omega_{j},T)=P^{i}_{S}(0)\Big[
g^{i \rightarrow j}_{\zeta_{1},\xi_{1}}(\tau_{i}-2,\omega_{i}-2,t-2\mid \tau_{j},\omega_{j})
-g^{i \rightarrow j}_{\zeta_{1},\xi_{1}}(\tau_{i}-2,\omega_{i}-2,t-1\mid \tau_{j},\omega_{j})
\\
\notag
&-g^{i \rightarrow j}_{\zeta_{2},\xi_{2}}(\tau_{i}-2,\omega_{i}-1,t-2\mid \tau_{j},\omega_{j})
+g^{i \rightarrow j}_{\zeta_{2},\xi_{2}}(\tau_{i}-2,\omega_{i}-1,t-1\mid \tau_{j},\omega_{j})
-g^{i \rightarrow j}_{\zeta_{3},\xi_{3}}(\tau_{i}-1,\omega_{i}-2,t-2\mid \tau_{j},\omega_{j})
\\
&+g^{i \rightarrow j}_{\zeta_{3},\xi_{3}}(\tau_{i}-1,\omega_{i}-2,t-1\mid \tau_{j},\omega_{j})
+g^{i \rightarrow j}_{\zeta_{4},\xi_{4}}(\tau_{i}-1,\omega_{i}-1,t-2\mid \tau_{j},\omega_{j})
-g^{i \rightarrow j}_{\zeta_{4},\xi_{4}}(\tau_{i}-1,\omega_{i}-1,t-1\mid \tau_{j},\omega_{j})
\Big]
\end{align}
for $1 \leq \tau_{i} \leq t-2$, $1 \leq \omega_{i} \leq t-1$, $0 \leq \tau_{j} \leq t-1$ and $1 \leq \omega_{j} \leq t$, where the functional $g^{i \rightarrow j}_{\zeta,\xi}(t_{1},t_{2},t_{3}\hspace{-0.05cm}\mid\hspace{-0.05cm}\tau_{j},\omega_{j})$ is defined as
\begin{equation}
g^{i \rightarrow j}_{\zeta,\xi}(t_{1},t_{2},t_{3}\hspace{-0.05cm}\mid\hspace{-0.05cm}\tau_{j},\omega_{j})=\zeta^{j \rightarrow i}(t_{1},t_{2},t_{3}\hspace{-0.05cm}\mid\hspace{-0.05cm}\tau_{j},\omega_{j})\hspace{-0.1cm}\prod_{k\in\partial i \backslash j}\hspace{-0.1cm}\xi^{k \rightarrow i}(t_{1},t_{2},t_{3}),
\label{eq:Kequations:g}
\end{equation}
and the $(\tau_{j},\omega_{j})$-dependent coefficients, characterizing the influence of node $j$ on the dynamics of $i$, are defined as follows for $t_{1}=t-2$ or $t_{1}=t-1$: 
\begin{align}
\notag
\zeta^{j \rightarrow i}_{1}(\tau_{i}-2,\omega_{i}-2,t_{1}\mid \tau_{j},\omega_{j})
=\prod_{t'=0}^{\tau_{i}-2}\left(1-\lambda_{ji}\mathds{1}[\tau_{j}\leq t']\mathds{1}[\omega_{j}\geq t'+1]\right)
&\prod_{t''=\tau_{i}}^{\omega_{i}-2}\left(1-\alpha_{ji}\mathds{1}[\tau_{j}\leq t'']\mathds{1}[\omega_{j}\geq t''+1]\right)
\\
&\times\prod_{t'''=\omega_{i}}^{t_{1}}\left(1-\beta_{ji}\mathds{1}[\omega_{j}\leq t''']
\right),
\\
\notag
\zeta^{j \rightarrow i}_{2}(\tau_{i}-2,\omega_{i}-1,t_{1}\mid \tau_{j},\omega_{j})
=\prod_{t'=0}^{\tau_{i}-2}\left(1-\lambda_{ji}\mathds{1}[\tau_{j}\leq t']\mathds{1}[\omega_{j}\geq t'+1]\right)
&\prod_{t''=\tau_{i}}^{\omega_{i}-1}\left(1-\alpha_{ji}\mathds{1}[\tau_{j}\leq t'']\mathds{1}[\omega_{j}\geq t''+1]\right)
\\
&\times\prod_{t'''=\omega_{i}}^{t_{1}}\left(1-\beta_{ji}\mathds{1}[\omega_{j}\leq t''']
\right),
\end{align}
\begin{align}
\notag
\zeta^{j \rightarrow i}_{3}(\tau_{i}-1,\omega_{i}-2,t_{1}\mid \tau_{j},\omega_{j})
=\prod_{t'=0}^{\tau_{i}-1}\left(1-\lambda_{ji}\mathds{1}[\tau_{j}\leq t']\mathds{1}[\omega_{j}\geq t'+1]\right)
&\prod_{t''=\tau_{i}}^{\omega_{i}-2}\left(1-\alpha_{ji}\mathds{1}[\tau_{j}\leq t'']\mathds{1}[\omega_{j}\geq t''+1]\right)
\\
&\times\prod_{t'''=\omega_{i}}^{t_{1}}\left(1-\beta_{ji}\mathds{1}[\omega_{j}\leq t''']
\right),
\\
\notag
\zeta^{j \rightarrow i}_{4}(\tau_{i}-1,\omega_{i}-1,t_{1}\mid \tau_{j},\omega_{j})
=\prod_{t'=0}^{\tau_{i}-1}\left(1-\lambda_{ji}\mathds{1}[\tau_{j}\leq t']\mathds{1}[\omega_{j}\geq t'+1]\right)
&\prod_{t''=\tau_{i}}^{\omega_{i}-1}\left(1-\alpha_{ji}\mathds{1}[\tau_{j}\leq t'']\mathds{1}[\omega_{j}\geq t''+1]\right)
\\
&\times\prod_{t'''=\omega_{i}}^{t_{1}}\left(1-\beta_{ji}\mathds{1}[\omega_{j}\leq t''']
\right).
\end{align}
The functions $\xi_{l}^{k \rightarrow i}$, $l=1\ldots4$, follow the evolution equations, similar to \eqref{eq:RSequations:chi1} and \eqref{eq:RSequations:chi2}:
\begin{align}
\xi_{1}^{k \rightarrow i}(\tau_{i}-2,\omega_{i}-2,t-1) =& \xi_{1}^{k \rightarrow i}(\tau_{i}-2,\omega_{i}-2,t-2)-\beta_{ki}\eta_{1}^{k \rightarrow i}(\tau_{i}-2,\omega_{i}-2,t-1),
\label{eq:Kequations:xi1}
\\
\xi_{2}^{k \rightarrow i}(\tau_{i}-2,\omega_{i}-1,t-1) =& \xi_{2}^{k \rightarrow i}(\tau_{i}-2,\omega_{i}-1,t-2)-\beta_{ki}\eta_{2}^{k \rightarrow i}(\tau_{i}-2,\omega_{i}-1,t-1),
\label{eq:Kequations:xi2}
\\
\xi_{3}^{k \rightarrow i}(\tau_{i}-1,\omega_{i}-2,t-1) =& \xi_{3}^{k \rightarrow i}(\tau_{i}-1,\omega_{i}-2,t-2)-\beta_{ki}\eta_{3}^{k \rightarrow i}(\tau_{i}-1,\omega_{i}-2,t-1),
\label{eq:Kequations:xi3}
\\
\xi_{4}^{k \rightarrow i}(\tau_{i}-1,\omega_{i}-1,t-1) =& \xi_{4}^{k \rightarrow i}(\tau_{i}-1,\omega_{i}-1,t-2)-\beta_{ki}\eta_{4}^{k \rightarrow i}(\tau_{i}-1,\omega_{i}-1,t-1).
\label{eq:Kequations:xi4}
\end{align}
It is straightforward to convince oneself that $\eta_{l}^{k \rightarrow i}$, $l=1\ldots4$, follow the equations of the type
\begin{align}
\notag
\eta_{1}^{k \rightarrow i}(\tau_{i}-2,\omega_{i}-2,t-1)&=(1-\beta_{ki}\mathds{1}_{\omega_{i} \neq t-1})\eta_{1}^{k \rightarrow i}(\tau_{i}-2,\omega_{i}-2,t-2)
\\
\notag
&+\sum_{\tau_{k}\leq t-2}\prod_{t'=0}^{\tau_{i}-2}\left(1-\lambda_{ki}\mathds{1}[\tau_{k}\leq t']\right)
\prod_{t''=\tau_{i}}^{t-2}\left(1-\alpha_{ki}\mathds{1}[\tau_{k}\leq t'']\right)
\mu^{k \rightarrow i}(\tau_{k},t-1\mid \tau_{i},\omega_{i})
\\
\notag
&-\sum_{\tau_{k} < \omega_{k} \leq t-2}\prod_{t'=0}^{\tau_{i}-2}\left(1-\lambda_{ki}\mathds{1}[\tau_{k}\leq t']\mathds{1}[\omega_{k} \geq t'+1]\right)
\prod_{t''=\tau_{i}}^{\omega_{i}-2}\left(1-\alpha_{ki}\mathds{1}[\tau_{k}\leq t'']\mathds{1}[\omega_{k} \geq t''+1]\right)
\\
&\times\prod_{t'''=\omega_{i}}^{t-2}\left(1-\beta_{ki}\mathds{1}[\omega_{k} \leq t''']\right)
m^{k \rightarrow i}(\tau_{k},\omega_{k},t-1\mid \tau_{i},\omega_{i},T).
\label{eq:Kequations:eta1}
\end{align}
The border conditions for $\omega_{i}=t-1$ are as follows:
\begin{align}
&\xi_{1}^{k \rightarrow i}(\tau_{i}-2,t-3,t-2)=\chi_{1}^{k \rightarrow i}(\tau_{i}-2,t-3),
\:\:\:\:\:
\eta_{1}^{k \rightarrow i}(\tau_{i}-2,t-3,t-2)=1-\psi_{1}^{k \rightarrow i}(\tau_{i}-2,t-2),
\\
&\xi_{2}^{k \rightarrow i}(\tau_{i}-2,t-2,t-2)=\chi_{1}^{k \rightarrow i}(\tau_{i}-2,t-2),
\:\:\:\:\:
\eta_{2}^{k \rightarrow i}(\tau_{i}-2,t-2,t-2)=1-(1-\alpha_{ki})\psi_{1}^{k \rightarrow i}(\tau_{i}-2,t-2),
\\
&\xi_{3}^{k \rightarrow i}(\tau_{i}-1,t-3,t-2)=\chi_{2}^{k \rightarrow i}(\tau_{i}-1,t-3),
\:\:\:\:\:
\eta_{3}^{k \rightarrow i}(\tau_{i}-1,t-3,t-2)=1-\psi_{2}^{k \rightarrow i}(\tau_{i}-1,t-2),
\\
&\xi_{4}^{k \rightarrow i}(\tau_{i}-1,t-2,t-2)=\chi_{2}^{k \rightarrow i}(\tau_{i}-1,t-2),
\:\:\:\:\:
\eta_{4}^{k \rightarrow i}(\tau_{i}-1,t-2,t-2)=1-(1-\alpha_{ki})\psi_{2}^{k \rightarrow i}(\tau_{i}-1,t-2),
\end{align}
and initial conditions are $\xi_{1}^{k \rightarrow i}(-2,-1,0)=\xi_{2}^{k \rightarrow i}(-2,0,0)=1$, $\eta_{1}^{k \rightarrow i}(-2,-1,1)=\eta_{2}^{k \rightarrow i}(-2,0,1)=\mu^{k \rightarrow i}(0,1 \mid 0,1)$.

\bibliography{library}
\end{document}